\documentclass[12pt]{article}
\usepackage{amssymb,cite,multirow}
\usepackage[dvips]{graphicx,epsfig}
\usepackage{graphics}

\def\DJ{{\hbox{D\kern-.8em\raise.15ex\hbox{--}\kern.35em}}}

\newcommand{\setall}{\setcounter{equation}{0}
        \setcounter{theorem}{0}}

\begin{document}

\rightline{hep-th/0701236}
\rightline{Bicocca-FT-07-03}
\rightline{SISSA 05/2007/EP}
\vskip 2cm


\centerline{\Huge Baryonic Generating Functions}
~\\

\renewcommand{\thefootnote}{\fnsymbol{footnote}}
\centerline{
 \bf Davide Forcella${}^{1}$\footnote{\tt forcella@sissa.it}, 
 \bf Amihay Hanany${}^{2}$\footnote{\tt ahanany@perimeterinstitute.ca}, 
 \bf Alberto Zaffaroni${}^{3}$\footnote{\tt alberto.zaffaroni@mib.infn.it} }
~\\
{\small
\begin{center}
${}^1$ International School for Advanced Studies (SISSA/ISAS) \\
and INFN-Sezione di Trieste,\\
via Beirut 2, I-34014, Trieste, Italy\\
~\\
${}^2$ Perimeter Institute for Theoretical Physics\\
31 Caroline Street North\\
Waterloo Ontario N2L 2Y5, Canada\\
~\\
${}^3$ Universit\`{a} di Milano-Bicocca\\
 and INFN, sezione di  Milano-Bicocca\\ 
Piazza della Scienza, 3; I-20126 Milano, Italy
\end{center}
}



\setcounter{footnote}{0}
\renewcommand{\thefootnote}{\arabic{footnote}}
\vskip 0.8in

\begin{abstract}

We show how it is possible to use the plethystic program in order to compute baryonic generating functions that count BPS operators in the chiral ring of quiver gauge theories living on the world volume of D branes probing a non compact CY manifold. Special attention is given to the conifold theory and the orbifold $ \frac{\mathbb{C}^2}{Z_2} \times \mathbb{C} $, where exact expressions for generating functions are given in detail. This paper solves a long standing problem for the combinatorics of quiver gauge theories with baryonic moduli spaces. It opens the way to a statistical analysis of quiver theories on baryonic branches. Surprisingly, the baryonic charge turns out to be the quantized K\"ahler modulus of the geometry.

\end{abstract}

\vfill
\newpage

\tableofcontents

\section{Introduction}
\setall
Counting problems in supersymmetric gauge theories reveal a rich structure of the chiral ring, its generators and their relations. Using the generating functions for counting BPS operators in the chiral ring, one can get further information about the dimension of the moduli space of vacua and the effective number of degrees of freedom in the system. 
The computation of such generating functions is generically a very hard problem and to date there is no clear way which demonstrates how to analyze such a problem. There are, however, special cases in which the problem simplifies considerably and one can get an exact answer for the generating function.

There is an amazing simplification of the problem of counting BPS operators in the chiral ring when the gauge theory is living on $N$ D brane in Type II superstring theory \cite{Benvenuti:2006qr,Butti:2006au,Feng:2007ur}. For such a case one is applying the plethystic program which is introduced in \cite{Benvenuti:2006qr} and discussed in further detail in \cite{Feng:2007ur}. The main point in this program is the use of the plethystic exponential and its inverse, the plethystic logarithm, in a way which allows the computation of finite $N$ generating functions in terms of only one quantity -- the generating function for a single D brane. The generating function for one D brane is calculated in turn by using geometric properties of the moduli space of vacua for this gauge theory, typically the moduli space being a Calabi-Yau (CY) manifold. In many cases the generating function for one D brane is computed even without the detailed knowledge of the quiver gauge theory and bypasses this data by just using the geometric properties of the moduli space.

The plethystic program brings about an important distinction between different types of moduli spaces -- we have three such types: 1. There are moduli spaces which are freely generated -- that is the set of generators obey no relations. 2. There are moduli spaces which are complete intersections -- that is there are generators that satisfy relations such that the number of relations plus the dimension of the moduli space is equal to the number of generators. 3. The rest are moduli spaces which are not complete intersections. The last class of models is by far the largest class. Most moduli spaces of vacua do not admit a simple algebraic description which truncates at some level. It is important, however, to single out and specify the moduli spaces which fall into the first two classes as they are substantially simpler to describe and to analyze.

Counting problems in supersymmetric gauge theories have been a subject of recent and not so recent interest. The first paper possibly dates back to 1998, \cite{Pouliot:1998yv} and is independently reproduced in \cite{Romelsberger:2005eg}. Other works include \cite{Kinney:2005ej, Biswas:2006tj, Mandal:2006tk,Martelli:2006vh,Basu:2006id}, this subject is of increased interest in recent times. Mesonic generating functions were given in \cite{Benvenuti:2006qr}, mixed branches were covered using ``surgery'' methods \cite{Hanany:2006uc} and baryonic generating functions for divisors in toric geometry were computed in \cite{Butti:2006au}. The combination of these problems brings about the next interesting problem -- that of counting baryonic operators on the moduli space.

Baryons have been considered in detail in supersymmetric gauge theories and were subject of intensive study in the context of the AdS/CFT correspondence \cite{Gubser:1998fp}. Of particular mention are the works by \cite{Berenstein:2002ke} were baryons were discussed in detail from a gauge theory point of view; \cite{Mikhailov:2000ya} with the relation of giant gravitons to holomorphic curves in the geometry; and \cite{Beasley:2002xv} with the quantization of the system, leading to the right way to treat counting of baryons in quiver gauge theories.

In this paper we progress one step further by computing generating functions for baryonic BPS operators. The crucial methods which are used are the successful combination of the plethystic program \cite{Benvenuti:2006qr, Feng:2007ur} together with the baryonic generating functions for divisors in the geometry \cite{Butti:2006au}. Together they form an exact result that computes BPS operators on baryonic branches.

The paper is organized as follows. In Section \ref{generalCY} we introduce the plethystic program for baryonic generating functions for theories with one baryonic charge. In Section \ref{conifold} we look in detail on baryonic generating functions for the conifold theory. We start by describing the case for $N=1$ D brane and discuss in detail the generating function for unfixed and for fixed baryonic charges. The analysis leads to an interesting interplay with Chebyshev polynomials and we spell out this correspondence. The relation to the geometry is elaborated and diagrammatic rules are composed for computing generating functions for $N=1$ D brane and fixed baryonic charge. The relation to tilings of the two dimensional plane is discussed and the lattice of BPS charges reveals itself beautifully as the (p,q) web of the toric diagram. We then proceed to discuss the general $N$ case and write the relations between the different generating functions, for fixed and unfixed D brane charge and for fixed and unfixed baryonic charges. The case of $N=2$ and small number of branes is computed and discussed in detail. Aspects of the plethystic exponential and the plethystic logarithm are then discussed in detail and lead to the understanding of the generators and relations for the chiral ring. We then turn to a toy model, termed ``half the conifold'' and demonstrate how the plethystic program for baryons computes the generating function exactly, leading to a freely generated moduli space for fixed number of branes, $N$. A comparison with the Molien invariant is then pursued. Another example called ``$\frac{3}{4}$ the conifold'' is discussed in detail. In Section \ref{secC2Z2} we compute the generating functions for the orbifold $\frac{\mathbb{C}^2}{Z_2}$ and finally we conclude.

\section{Generating Functions for CY Manifolds with one Baryonic Charge}
\label{generalCY}
\setall

In this section we will give general prescriptions on the computation of generating functions for BPS operators in the chiral ring of a supersymmetric gauge theory that lives on a D brane which probes a generic non-compact CY manifold which has a single baryonic charge. A class of such manifolds includes the $Y^{p,q}$ theories \cite{gauntlett}, or more generally it includes the $L^{abc}$ theories \cite{CLPP,kru2,noi}. A challenge problem which remained elusive up to now was to compute the generating functions for baryonic operators. This paper will be devoted to the study of this problem, with success for a selected set of theories. The problem of computing generating functions for a general CY manifold is still very difficult to solve but with the methods of this section we will be able to reduce the problem of computing the generating function for a generic number of D branes, $N$ and generic baryonic number $B$, to a much simpler problem. This reduction is done using the plethystic program \cite{Benvenuti:2006qr} (See further details in \cite{Feng:2007ur}).

Recall that in the case of baryon number $B=0$, namely for mesonic generating functions \cite{Benvenuti:2006qr}, the knowledge of the generating function for $N=1$ is enough to compute the generating function for any $N$. This is essentially due to the fact that the operators for finite $N$ are symmetric functions of the operators for $N=1$ and this is precisely the role which is played by the plethystic exponential -- to take a generating function for a set of operators and count all possible symmetric functions of it.

For the present case, in which the baryon number is non-zero, it turns out that the procedure is not too different than the mesonic case. One needs to have the knowledge of a single generating function, $g_{1,B}$, for one D brane, $N=1$ and for a fixed baryon number $B$, and this information is enough to generate all generating functions for any number of D branes and for a fixed baryonic number \cite{Butti:2006au}.

Given a ${\cal N}=1$ supersymmetric gauge theory with a collection of $U(1)$ global symmetries, $\prod_{i=1}^G U(1)_i$, we will have a set of $G$ chemical potentials $\{t_i\}_{i=1}^G$. The generating function for a gauge theory living on a D brane probing a generic non-compact CY manifold is going to depend on the set of parameters, $t_i$. There is always at least one such $U(1)$ global symmetry and one such chemical potential $t$, corresponding to the $U(1)_R$ symmetry.
For a given CY manifold we will denote the generating function for a fixed number of D branes, $N$ and a fixed baryonic charge $N B$ by $g_{N,N B}(\{t_i\}; CY).$

Let us introduce a chemical potential $\nu$ for the number of D branes. Then the generating function for any number of D branes and for a fixed baryonic charge $B$ is given by

\begin{eqnarray}
\nonumber
g_B(\nu; \{t_i\}; CY) &=& \hbox{PE$_{\nu}$}[g_{1,B}(\{t_i\}; CY)] \equiv \exp\biggl(\sum_{k=1}^\infty \frac{\nu^k}{k}g_{1,B}(\{t_i^k\}; CY) \biggr)\\
&=& \sum_{N=0}^\infty \nu^N g_{N,N B}(\{t_i\}; CY) .
\label{g1plet}
\end{eqnarray}

This formula can be inverted to give the generating function for fixed baryonic charge and fixed number of branes, in case the generating function for fixed baryonic charge is known,

\begin{equation}
\nonumber
g_{N,N B}(\{t_i\}; CY) = \frac{1}{2\pi i} \oint \frac{d\nu}{\nu^{N+1}} g_B(\nu; \{t_i\}; CY) .
\end{equation}

We can further introduce a chemical potential $b$ for the baryon number and sum over all possible baryon numbers go get the generating function for an unfixed baryonic number,

\begin{eqnarray}
g(\nu; \{t_i\}; CY) = \sum_{B=-\infty}^\infty b^{N B} \exp\biggl(\sum_{k=1}^\infty \frac{\nu^k}{k}g_{1,B}(\{t_i^k\}; CY) \biggr) .
\end{eqnarray}

This formula is very schematic, however sufficient to the purposes
of this paper. The correct general formula would accommodate
multiple baryonic charges, not necessarily running over all integers, and
multiplicities for the components $g_B$. 
From the previous formula we can extract the generating function for a fixed number of D branes $N$,

\begin{eqnarray}
g(\nu; \{t_i\}; CY) = \sum_{N=0}^\infty \nu^N g_N(\{t_i\}; CY).
\end{eqnarray}

This set of equations form the basis of this paper and allow for the computation of $g_{N,N B}$ for any $N$ and $B$, once $g_1$ is known. As a result, the problem of computing generating functions for baryonic operators greatly simplifies and amounts to figuring out the much simpler case of one D brane, $g_1$.

The generating function $g_{1,B}$ can be given a geometric interpretation. As $g_{1,0}$ is the
character for holomorphic functions on the CY manifold \cite{Martelli:2006yb}, $g_{1,B}$ is the
character for holomorphic sections of suitable line bundles 
\cite{Butti:2006au}. 
This is essentially due to the fact that we can use holomorphic surfaces to parameterize supersymmetric configurations of D3 branes wrapped on non trivial cycles on the CY horizon \cite{Mikhailov:2000ya,Beasley:2002xv}. In this construction the baryonic charge in gauge theory is related to the homology of the cycles
in geometry and $g_{1,B}$ gets contribution from surfaces 
(zero loci of sections) belonging to a given homology class. 
Formula (\ref{g1plet}) can be alternatively interpreted as giving 
the BPS Hilbert space of states with baryonic charge $B$ and number of branes $N$ as the geometric
quantization of the classical configuration space of supersymmetric D3
branes wrapped on cycles of homology $B$. The explicit construction and
methods for computing  $g_{1,B}$ have been given in \cite{Butti:2006au}.

To demonstrate this method of computation we will now turn to few simple examples in which one can compute the generating functions and perform some consistency checks of this proposal.

\section{Baryonic Generating Functions for the Conifold}
\label{conifold}
\setall

The gauge theory on the conifold has a global symmetry $SU(2)_1\times SU(2)_2\times U(1)_R\times U(1)_B$. It has 4 basic fields $A_{1,2}$ and $B_{1,2}$ that transform under these symmetries according to the following table:

\begin{table}[htdp]
\caption{Global charges for the basic fields of the quiver gauge theory on the D brane probing the Conifold.}
\begin{center}
\begin{tabular}{|c||c c|c c|c|c||c|}
\hline
 & \multicolumn{2} {c|} {$SU(2)_1$} & \multicolumn{2} {c|} {$SU(2)_2$} & $U(1)_R$ & $U(1)_B$ & monomial\\ \hline
\cline{2-5}
& $j_1$ & $m_1$ & $j_2$ & $m_2$&&&\\ \hline \hline
$A_1$ & $\frac{1}{2}$ & $+\frac{1}{2}$ & 0 & 0 & $\frac{1}{2}$ & 1& $t_1 x$\\ \hline
$A_2$ & $\frac{1}{2}$ & $-\frac{1}{2}$ & 0 &0 & $\frac{1}{2}$ & 1& $\frac{t_1}{x}$ \\ \hline
$B_1$ & 0 & 0 & $\frac{1}{2}$ & $+\frac{1}{2}$ & $\frac{1}{2}$ & -1& $t_2 y$ \\ \hline
$B_2$ & 0 & 0 & $\frac{1}{2}$ & $-\frac{1}{2}$ & $\frac{1}{2}$ & -1& $\frac{t_2}{y}$ \\ \hline
\end{tabular}
\end{center}
\label{globalconifold}
\end{table}

The last column represents the corresponding monomial in the generating function for BPS operators in the chiral ring. $t_1$ is the chemical potential for the number of $A$ fields, $t_2$ is the chemical potential for the number of $B$ fields, $x$ is the chemical potential for the Cartan generator of $SU(2)_1$ and $y$ is the chemical potential for the Cartan generator of $SU(2)_2$. A generic operator that carries a spin $j_1$ with weight $m_1$ under $SU(2)_1$, spin $j_2$ with weight $m_2$ under $SU(2)_2$ has $2j_1$ $A$'s, $2j_2$ $B$'s and therefore will be represented by the monomial

\begin{equation}
t_1^{2j_1} x^{2m_1} t_2^{2j_2} y^{2m_2}.
\label{monom}
\end{equation}

We can also keep track of the R-charge and the baryonic charge $B$ by introducing chemical potentials $t$ and $b$, respectively. With this notation we have $t_1=t b$ and $t_2 = \frac{t}{b}$, and a generic operator represented by the monomial (\ref{monom}) is given by

\begin{equation}
t^{2j_1+2j_2} x^{2m_1} b^{2j_1-2j_2} y^{2m_2},
\end{equation}
and carries $R=j_1+j_2$ and $B=2(j_1-j_2)$.
With this notation we can proceed and write generating functions which count BPS operators in the chiral ring of the conifold theory.

\subsection{$N=1$ for the Conifold}
\label{N1}

Since the superpotential for the $N=1$ theory is $W=0$, it is natural to expect that the generating function is freely generated by the 4 basic fields of the conifold gauge theory and it takes the form

\begin{equation}
g_1(t_1,t_2,x,y; {\cal C}) = \frac{1}{(1-t_1 x) (1-\frac{t_1}{x})(1-t_2 y) (1-\frac{t_2}{y})}.
\label{g1coni}
\end{equation}

\begin{figure}[h!!!]
\begin{center}
\includegraphics[scale=0.55]{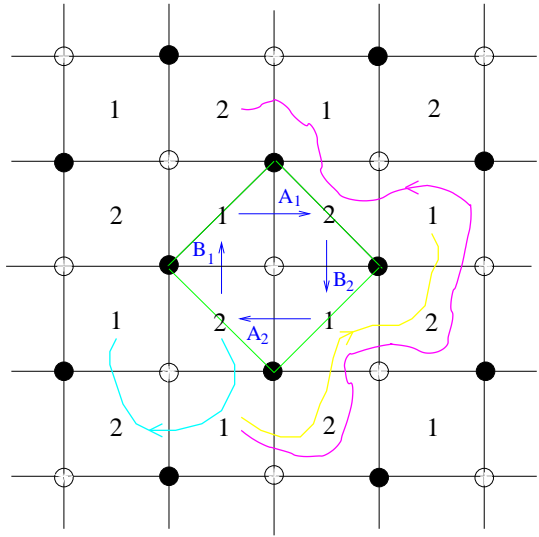} 
\caption{The dimer representation for the conifold. The fundamental domain is depicted in green. We have drawn some paths representing BPS operators in the chiral ring in the case $N=1$: in cyan $B_2A_2B_1$; in yellow $A_1B_1A_1B_1$, in magenta $A_1B_1A_1B_1A_2B_1A_2$.}
\label{coni}
\end{center}
\end{figure}

This generating function has a simple interpretation in the dimer representation \cite{Hanany:2005ve} for the conifold (see Figure \ref{coni}). Recall that the tiling for the conifold gets the form of a chessboard array \cite{Franco:2005rj} with two tiles in the fundamental domain and with horizontal arrows $A_{1,2}$ going right and left respectively, and vertical arrows $B_{1,2}$ going up and down respectively. A generic BPS operator in the chiral ring for $N=1$ will be represented by an open path or possibly by many open paths in the tiling and the function (\ref{g1coni}) represents all such open paths.

The dimension of the moduli space for the $N=1$ theory is 4, as can be seen from the behavior near $t_i =1$, where one finds the generating function for $\mathbb{C}^4$: 

\begin{equation}
g_1(t_1=t,t_2=t,x=1,y=1; {\cal C}) = \frac{1}{(1-t)^4}.
\end{equation}

This is natural to expect, as this theory has 4 fundamental fields, a global symmetry of rank 4, and a vanishing superpotential, leading to a freely generated chiral ring. We will see other examples in which these conditions are not met and, as a result, the behavior is different.

It is useful to rewrite equation (\ref{g1coni}) in terms of the baryonic and R-charge chemical potentials.

\begin{equation}
g_1(t,b,x,y; {\cal C}) = \frac{1}{(1-t b x) (1-\frac{t b}{x})(1- \frac {t y}{b}) (1-\frac{t}{b y})},
\label{g1conitbxy}
\end{equation}
and to expand $g_1$ in a  ``generalized'' Laurent expansion

\begin{equation}
g_1(t,b,x,y; {\cal C}) = \sum_{B=-\infty}^\infty b^B g_{1,B}(t,x,y; {\cal C}),
\end{equation}
where $g_{1,B}(t,x,y; {\cal C})$ is the generating function for BPS operators at fixed number of branes with $N=1$ and fixed baryonic charge $B$. It can be computed using the inversion formula

\begin{equation}
g_{1,B}(t,x,y; {\cal C}) = \frac{1}{2\pi i} \oint \frac {db} {b^{B+1}} g_1(t,b,x,y; {\cal C})\equiv \frac{1}{2\pi i} \oint db I ,
\label{res1}
\end{equation}
with a careful evaluation of the contour integral for positive and negative values of the baryonic charge $B$. We have denoted for simplicity the integrand to be $I$. For $B\ge0$ the contribution of the contour integral comes from the positive powers of the poles for $b$ and we end up evaluating the two residues 

\begin{equation}
B\ge0 : - res|_{b=\frac{1}{t x}} I - res|_{b=\frac{x}{t}} I, \nonumber
\label{res2}
\end{equation}

For $B\le0$ the contribution of the contour integral comes from the negative powers of the poles for $b$ and we end up evaluating the two residues 

\begin{equation}
B\le0 : res|_{b=\frac{t}{y}} I + res|_{b= t y} I .
\label{res3}
\end{equation}

Collecting these expressions together we get the generating functions for fixed baryonic number

\begin{eqnarray}
g_{1,B\ge0}(t,x,y; {\cal C}) &=& \frac{t^B x^{B} } { (1 - \frac{1}{x^2}) (1-t^2 x y)  (1-\frac{t^2 x}{y}) }+ \frac{t^B x^{-B}} { (1 - x^2) (1-\frac{t^2 y} {x})  (1-\frac{t^2}{x y}) } , \nonumber \\ \nonumber
g_{1,B\le0}(t,x,y; {\cal C}) &=& \frac{t^{-B} y^{-B} } { (1-\frac{1}{y^2}) (1-t^2 x y)  (1-\frac{t^2 y}{x}) }+ \frac{t^{-B} y^{B}} { (1 - y^2) (1-\frac{t^2 x} {y})  (1-\frac{t^2}{y x}) } .\\
\label{rescon}
\end{eqnarray}

Indeed each of these equations reflects the Weyl symmetry of each $SU(2)$ global symmetry, acting as $x\leftrightarrow\frac{1}{x}$, $y\leftrightarrow\frac{1}{y}$, respectively. Furthermore, under the map of $B\leftrightarrow-B$ and $x\leftrightarrow y$ these equations are invariant, reflecting the fact that the two $SU(2)$ global symmetries are exchanged when the baryon number reverses sign.

\begin{figure}[h!!!]
\begin{center}
\includegraphics[scale=0.65]{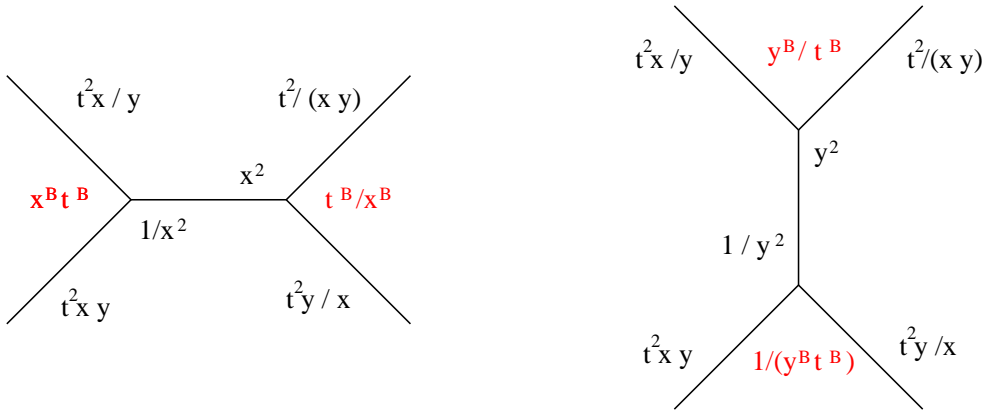} 
\caption{The $(p,q)$ webs for the two possible resolution of the conifold. The left figure corresponds to positive baryon number, $B>0$, while the right figure corresponds to negative baryon number, $B<0$.
The vertices correspond to the fixed points of the torus action on the 
resolution.}
\label{conifoldlocalization2}
\end{center}
\end{figure}
The equations for $g_{1,B}$ take the general form of equation 6.8 in \cite{Butti:2006au}. It is therefore suggestive that the formulas for $g_{1,B}$ will always be of this form and will come from localization, giving rise to contributions from the vertices of the $(p,q)$ web which are the fixed points of the torus action. Figure \ref{conifoldlocalization2} shows the contribution of each leg and each vertex in the $(p,q)$ web of the conifold theory. The left (right) vertex gives rise to the first (second) contribution for $g_{1,B\ge0}$, respectively, while the bottom (top) vertex gives rise to the first (second) contribution for $g_{1,B\le0}$, respectively. A more detailed discussion of the localization formula
for the conifold is given in Section \ref{geometricamusement}.

We can ignore the $SU(2)$ quantum numbers in Equations (\ref{rescon}) by setting $x=y=1$ and get

\begin{equation}
g_{1,B}(t,1,1; {\cal C}) = \frac{t^{ |B| } (1+t^2)} { (1- t^2)^3}+ \frac{ |B| t^{ |B| } } { (1- t^2 )^2 } ,
\label{rescon11}
\end{equation}
which indeed for the mesonic generating function, $B=0$, coincides with Equation (4.5) of \cite{Benvenuti:2006qr} and generalizes it to any baryonic number.

\subsubsection{Chebyshev Gymnastics}

It is useful to note the natural appearance of Chebyshev polynomials due to the fact that the global symmetry involves $SU(2)$ factors. In fact since we have two such factors we should expect two independent series of Chebyshev polynomials.

Let us recall equation (\ref{g1coni}), the generating function for one D brane, $N=1$,  on the conifold, ${\cal C}$,

\begin{equation}
g_1(t_1,t_2,x,y; {\cal C}) = \frac{1}{(1-t_1 x) (1-\frac{t_1}{x})(1-t_2 y) (1-\frac{t_2}{y})},
\end{equation}
and set $x=e^{i\theta_1}$, $y=e^{i\theta_2}$. Using the generating identity for $U_n(\cos\theta)$, the Chebyshev polynomials of the second kind

\begin{equation}
\frac{1}{1-2 t \cos\theta +t^2} = \sum_{n=0}^\infty U_{n} (\cos\theta) t^n,
\end{equation}
we find that 

\begin{eqnarray}
g_1(t_1,t_2,x,y; {\cal C}) &=& \sum_{n_1=0}^\infty \sum_{n_2=0}^\infty U_{n_1} (\cos\theta_1) t_1^{n_1} U_{n_2} (\cos\theta_2) t_2^{n_2} \\
&=& \sum_{j_1} \sum_{j_2} t_1^{2j_1} t_2^{2j_2} U_{2j_1}(\cos\theta_1) U_{2j_2} (\cos\theta_2) ,
\end{eqnarray}
where $j_1$ and $j_2$ run over all spin representations of $SU(2)$.
We can restore the baryonic charge dependence by using the chemical potentials $t$ and $b$ and get

\begin{eqnarray}
g_1(t_1,t_2,x,y; {\cal C})
&=& \sum_{n_1=0}^\infty \sum_{n_2=0}^\infty t^{n_1+n_2} b^{n_1-n_2} U_{n_1}(\cos\theta_1) U_{n_2} (\cos\theta_2) \\
&=& \sum_{B=-\infty}^\infty b^{B} \sum_{n=0}^\infty t^{2n+B} U_{n+B}(\cos\theta_1) U_{n} (\cos\theta_2) ,
\end{eqnarray}
where the second sum is derived by using the substitution $n_1=n+B, n_2=n$.
The generating functions for fixed baryonic charge then take the form

\begin{eqnarray}
\nonumber
g_{1,B\ge0}(t,x,y; {\cal C})
&=& \sum_{n=0}^\infty t^{2n+B} U_{n+B} (\cos \theta_1) U_{n} (\cos \theta_2) \\ \nonumber
&=& \frac{t^B x^{B+1} } { (x-\frac{1}{x}) (1-t^2 x y)  (1-\frac{t^2 x}{y}) }+ \frac{t^B x^{-B-1}} { (\frac{1}{x}-x) (1-\frac{t^2 y} {x})  (1-\frac{t^2}{x y}) } , \\
g_{1,B\le0}(t,x,y; {\cal C})
\label{chebyid}
&=& \sum_{n=0}^\infty t^{2n+|B|} U_n (\cos \theta_1) U_{n+|B|} (\cos \theta_2) \\ \nonumber
&=& \frac{t^{|B|} y^{|B|+1} } { (y-\frac{1}{y}) (1-t^2 x y)  (1-\frac{t^2 y}{x}) }+ \frac{t^{|B|} y^{-|B|-1}} { (\frac{1}{y}-y) (1-\frac{t^2 x} {y})  (1-\frac{t^2}{y x}) } ,
\end{eqnarray}
where we have made use of the residue formulas, (\ref{rescon}). Equations (\ref{chebyid}) form some Chebyshev identities with two variables where the computation is done by two different methods. For the special case $B=0$ we find another identity for Chebyshev polynomials

\begin{eqnarray}
\nonumber
g_{1,B=0}(t,x,y; {\cal C})
&=& \sum_{n=0}^\infty t^{2n} U_{n} (\cos \theta_1) U_{n} (\cos \theta_2) \\ \nonumber
&=& \frac{x} { (x-\frac{1}{x}) (1-t^2 x y)  (1-\frac{t^2 x}{y}) }+ \frac{x^{-1}} { (\frac{1}{x}-x) (1-\frac{t^2 y} {x})  (1-\frac{t^2}{x y}) } \\ \nonumber
&=& \frac{y} { (y-\frac{1}{y}) (1-t^2 x y)  (1-\frac{t^2 y}{x}) }+ \frac{y^{-1}} { (\frac{1}{y}-y) (1-\frac{t^2 x} {y})  (1-\frac{t^2}{y x}) } \\
&=& \frac{1-t^4} {(1-t^2 x y) (1-\frac{t^2 y}{x}) (1-\frac{t^2 x} {y}) (1-\frac{t^2}{y x}) } .
\label{chebymesonid}
\end{eqnarray}

\subsubsection{Geometric Amusement}
\label{geometricamusement}

It is interesting to re-interpret the $N=1$ generating function that we have found in Equations (\ref{g1coni}) and (\ref{rescon}) from a geometrical point of view. We can describe the conifold in terms of homogeneous coordinates,

\begin{equation}
\{ x_1\sim x_1 \mu\,  , \quad x_2 \sim \frac{x_2}{\mu}\, ,\quad x_3\sim x_3 \mu\, , \quad x_4\sim \frac{x_4}{\mu} \}\qquad\qquad \mu\in \mathbb{C}^*
\label{eqcon}
\end{equation}

\begin{figure}[h!!!!!]
\begin{center}
\includegraphics[scale=0.5]{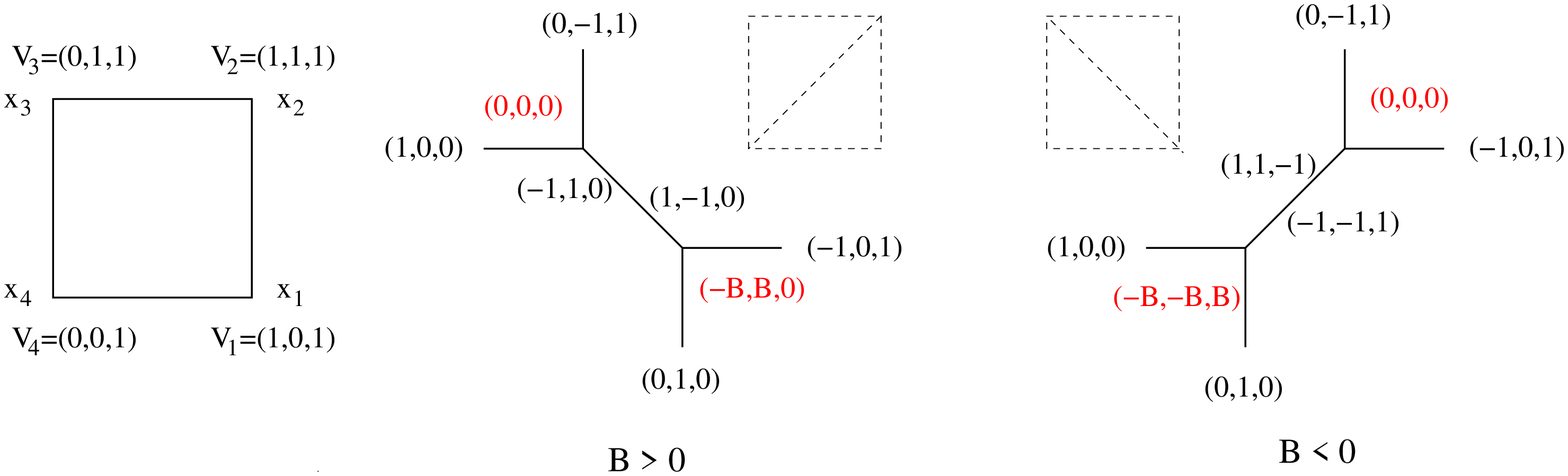} 
\caption{Localization data for the $N=1$ baryonic partition functions. The
vertices $V_i$ are in correspondence with homogeneous coordinates $x_i$ and
with a basis of  divisors $D_i$. Two different resolutions, related by
a flop, should be used for positive and negative $B$, respectively. 
Each resolution has two fixed points,
corresponding to the vertices of the $(p,q)$ webs; the weights $m^{(I)}_i,\, i=1,2,3$
and $m^{(I)}_B$ at the fixed points are indicated in black and red, respectively.}
\label{co}
\end{center}
\end{figure}

Homogeneous coordinates can be introduced for all toric varieties and 
generalize the familiar homogeneous coordinates for projective spaces
$\mathbb{P}^k$. The general rule is that there is a homogeneous coordinate
for each vertex $V_i\, i=1,\ldots,d$ in the toric diagram. These are
subject to the rescaling $x_i\sim \mu_i x_i$ for all $\mu_i$ that
satisfy
\begin{equation}
\prod_{i=1}^d \mu_i^{\left<e_k,V_i\right>} =1,\qquad k=1,2,3,
\label{bartoric}
\end{equation}
for the three basis vectors $e_k, k=1\ldots3$ of $\mathbb{Z}^3$.
See Figure \ref{co} for the conifold case. 

The conifold can be described as the algebraic quotient of $\mathbb{C}^4$ 
by the $\mathbb{C}^*$ action given in equation (\ref{eqcon}).
In this description all the four abelian symmetries of the
gauge theory are manifest as the four isometries of $\mathbb{C}^4$, which 
descend to the R symmetry, the two flavor symmetries and one baryonic symmetry of the 
theory on the conifold. In $\mathbb{C}^4$ the symmetries act 
directly on the homogeneous coordinates. We will use notations where
$x_i$ denotes both the homogeneous coordinates and the chemical potentials
for the standard basis of the four $U(1)$ symmetries of $\mathbb{C}^4$.    
In the notation of the previous section, we can set the correspondence

\begin{equation}
(x_1,\, x_2,\, x_3,\, x_4)\, =\, (t b x, \frac{t y}{b}, \frac{t b}{x}, \frac{t}{b y}) .
\end{equation}

Note in particular that the rescaling symmetry of the $x_i$, by which
we mod out to obtain the conifold, can be identified with the baryonic 
symmetry. More generally, a CY with a toric diagram with $d$ vertices
is the algebraic quotient of $\mathbb{C}^d$ by the $d-3$ $\mathbb{C}^*$ actions
given by the $\mu_i$ obeying (\ref{bartoric}) which correspond to the $d-3$
baryonic symmetry of the dual gauge theory.

In the correspondence between tiling and toric cones, 
homogeneous coordinates play an important role in identifying the charges
of the elementary fields in the gauge theory; in the simple case of the
conifold there is just a one-to-one correspondence between the four
homogeneous coordinates $(x_1,x_2,x_3,x_4)$ and the four elementary
fields $(A_1,B_1,A_2,B_2)$. 

Now we are ready for discussing the geometric interpretation of $g_{1,B}$.
The $N=1$ generating function $g_1(b,t,x,y; {\cal C})$ is just
the generating function of homogeneous monomials $x^\alpha=\prod_{i=1}^4 x_i^{\alpha_i}$ on the conifold. We can assign a degree $B$ to a monomial by
looking at the scaling behavior under the $\mathbb{C}^*$ action in equation
(\ref{eqcon}): $x^\alpha\rightarrow \lambda^B x^\alpha$.
The generating functions $g_{1,B}(t,x,y; {\cal C})$ count the subset
of monomials of degree $B$. This agrees with our previous definition
of $g_{1,B}(t,x,y; {\cal C})$ as the generating function for operators
of baryonic charge $B$ since the elementary fields are in correspondence with
the homogeneous coordinates and the $\mathbb{C}^*$ action 
coincides with the baryonic symmetry. 

Homogeneous polynomials
 are not functions on the conifold in a strict sense, as they transform non trivially under a rescaling of the $x_i$, but are rather sections of a line bundle.
Line bundles on the conifold are labeled by an integer B. For simplicity,
we denote them as ${\cal O}(B)$. The holomorphic sections of 
${\cal O}(B)$ are just the polynomials of degree $B$. 
The construction exactly  parallels the familiar case of $\mathbb{P}^k$ where 
the homogeneous polynomials of degree $n$ are the sections of
the line bundle ${\cal O}(n)$. In toric geometry, sections of line bundles, when arranged according to
their charge under the torus action, fill a convex polytope in $\mathbb{Z}^3$.
For each integer point $P$ in the polytope there is exactly one section with charges given by the integer entries of the point $P\in\mathbb{Z}^3$. For the
conifold, these polytopes have the shape of  integral conical pyramids and
will be discussed in details in the next Section.

It is a general result of toric geometry
that the homogeneous coordinate ring of a toric variety 
decomposes as a sum over non-trivial
line bundles \cite{Cox:1993fz}, which is equivalent to the statement that
polynomials can be graded by their degree. In the case of the conifold we have

\begin{equation}
\mathbb{C}[x_1,x_2,x_3,x_4] = \sum_{B=-\infty}^{\infty} {\rm H}^0 ({\cal C}, {\cal O}(B)) .
\end{equation}

The equation
\begin{equation}
g_1(t,b,x,y; {\cal C}) = \sum_{B=-\infty}^\infty b^B g_{1,B}(t,x,y; {\cal C}),
\end{equation}
translates the previous mathematical identity at the level of generating
functions. In this context 
$g_{1,B}(t,x,y; {\cal C})$ is interpreted as a
 character under the action of the three abelian isometries of the conifold
with chemical potential $x,y,t$,
\begin{equation}
g_{1,B}(t,x,y; {\cal C}) = {\mbox Ch}\, {\rm H}^0({\cal C}, {\cal O}(B))\equiv
{\rm Tr}\{x,y,t| {\rm H}^0({\cal C}, {\cal O}(B))\} .
\label{charac}
\end{equation}

The most efficient way of computing the $g_{1,B}(t,x,y; {\cal C})$ is to expand
the generating function of the coordinate ring $g_1$ in a generalized 
Laurent series, but it is interesting to observe that the character (\ref{charac}) can be alternatively computed using the index theorem.

The way of doing this computation is explained in detail in \cite{Butti:2006au} and expresses
the result as a sum over the fixed points $P_I$ of the torus $T^3$ action in a
smooth resolution of the conifold

\begin{equation}
g_{1,B}(q;  {\cal C}) = q^{n_B}\sum_{P_I} \frac{q^{m^{(I)}_B}}{\prod_{i=1}^3 (1-q^{m^{(I)}_i})} ,
\label{loc}
\end{equation}
where the index $I$ denotes the set of isolated fixed points
and the four vectors
$m^{(I)}_i,\, i=1,2,3$, $m^{(I)}_B$ in $\mathbb{Z}^3$ are the weights of the linearized action of $T^3$ on ${\cal C}$  and the fiber of the line bundle, respectively.

We refer to Section 6 of \cite{Butti:2006au} for the details of 
the computation and we just explain the meaning of the various terms
in formula (\ref{loc}). First of all, we notice that to perform a 
computation using the index theorem we need to choose a correctly normalized
basis $q\equiv(q_1,q_2,q_3)$ for the $T^3$ torus action on the manifold.
This implies taking square roots of the variables $(x,y,t)$ (that have
been chosen for notational simplicity). The relation between variables
is easy computed by using
\begin{equation}
q_k=\prod_{i=1}^4 x_i^{\langle e_k,V_i\rangle}\,\qquad\qquad k=1,2,3, 
\label{rel}
\end{equation} 
where $e_k$ are the basis vectors of $\mathbb{Z}^3$ and $V_i$ the vertices of
the toric diagram (notice that all dependence on baryonic charges drops from
the right hand side by equation (\ref{bartoric})).
We thus have

\begin{eqnarray}
\nonumber
q_1&=&x_1 x_2 = t^2 x y,\\ \nonumber
q_2&=&x_2 x_3 = \frac{t^2 y}{x},\\ 
q_3&=&x_1 x_2 x_3 x_4 = t^4 .
\label{GLr}
\end{eqnarray}
 
The two possible resolutions of the conifold are shown in  Figure \ref{co}.
The fixed points of the 
torus action are in correspondence with the vertices of the $(p,q)$ web
or, equivalently with the triangles in the subdivision of the toric diagram. 
The vectors  $m^{(I)}_i,\, i=1,2,3$ in the denominator of
formula (\ref{loc}) are computed as the three primitive inward normal vectors
of the cone $\sigma_I$ in $\mathbb{Z}^3$ made with the three vertices $V_{i}$ of
the I-th triangle. For computing  the character $g_{1,0}$ of holomorphic 
functions on the conifold we use formula (\ref{loc}) with $m_B^{(I)}=n_B=0$ \cite{Martelli:2006yb}; both resolutions can be used and give the same result.   
For computing the character $g_{1,B}$ for sections of a line bundle of degree
$B$ we need to choose a convenient resolution and compute the vectors 
$m_B^{(I)}$ and $n_B$. This can be done as follows. A generic line bundle
can be associated with a linear combination $\sum_{i=1}^d c_i D_i$ of the basic
divisors. There is a basic divisor $D_i$ for each vertex and we can assign
to it the same charges of the corresponding $x_i$; two divisors are linearly
equivalent if and only if they have the same baryonic charge. Given a divisor
$\sum_{i=1}^d c_i D_i$ of degree $B$ we can assign numbers $c_i$ to the
vertices $V_i$ of the toric diagram. Each fixed point $I$ determines
a vector $m_B^{(I)}$ as the integer solution of the linear system of three
equations   

\begin{equation}\label{charge}
\langle m_B^{(I)}, V_i \rangle = - c_i, \,\,\, V_i\in \sigma_I,  
\end{equation}
where the $V_i$ are the vertices of the $I$-th triangle.
A resolution gives the correct result for $g_{1,B}$ whenever all the vectors $m_B^{(I)}$ satisfy the convexity condition
$\langle m_B^{(I)}, V_i \rangle \ge - c_i$ for the vertices $V_i$ not belonging to
the $I$-th triangle.
Finally, the prefactor $q^{n_B}$ in formula (\ref{loc}) is just given
by the $T^3$ charge of the monomial $\prod_{i=1}^d x_i^{c_i}$.

In the case at hand, the partition functions $g_{1,B}$ can be
computed using the divisor $B D_1$ and the resolution
on the left in Figure (\ref{co}) for $B>0$
and the divisor $|B| D_4$ and the resolution on the right 
for $B<0$. It is interesting to note that in passing from positive to 
negative baryonic charges we need to perform a flop on the resolved 
conifold. We will
say more about the flop transition in the next Section. 
The weights are reported in Figure \ref{co}. Formula (\ref{loc}) gives

\begin{eqnarray}
g_{1,B\ge0}(q; {\cal C}) &=& {q_3^{\frac{B}{4}} (\frac{q_1}{q_2})^{\frac{B}{2}}}
\left(\frac{1 } { (1 - q_1) (1-\frac{q_3}{q_2})  (1-\frac{q_2}{q_1}) }+ \frac{( \frac{q_2}{q_1})^B} { (1 - q_2) (1-\frac{q_3} {q_1})  (1-\frac{q_1}{q_2}) }\right) , \nonumber \\ \nonumber
g_{1,B\le0}(q; {\cal C}) &=& \frac{(q_1 q_2)^{\frac{B}{2}}}{q_3^{\frac{3B}{4}}}
\left( \frac{(\frac{q_3}{q_1 q_2 })^B } { (1-q_1) (1-q_2)  (1-\frac{q_3}{q_1 q_2}) }+ \frac{1} { (1-\frac{q_3}{q_2}) (1-\frac{q_3} {q_1})  (1-\frac{q_1 q_2}{q_3}) }\right ) .
\end{eqnarray}

This formula, when expressed in terms of $(x,y,t)$, using Equation (\ref{GLr}), becomes

\begin{eqnarray}
g_{1,B\ge0}(t,x,y; {\cal C}) &=& \frac{t^B x^{B} } { (1 - \frac{1}{x^2}) (1-t^2 x y)  (1-\frac{t^2 x}{y}) }+ \frac{t^B x^{-B}} { (1 - x^2) (1-\frac{t^2 y} {x})  (1-\frac{t^2}{x y}) } , \nonumber \\ \nonumber
g_{1,B\le0}(t,x,y; {\cal C}) &=& \frac{t^{-B} y^{-B} } { (1-\frac{1}{y^2}) (1-t^2 x y)  (1-\frac{t^2 y}{x}) }+ \frac{t^{-B} y^{B}} { (1-y^2) (1-\frac{t^2 x} {y})  (1-\frac{t^2}{y x}) } .
\end{eqnarray}
which coincides with the result obtained in the previous Sections. 
It is amusing that the same formula can be interpreted as a result of a direct 
computation, of Chebyshev identities and of a localization theorem.

The extension of the previous discussion to a generic toric cone is discussed
in \cite{Butti:2006au}. It is a general fact that inequivalent line bundles
on toric cones are labelled by baryonic charges and that the homogeneous
coordinate ring decomposes as direct sum of all inequivalent line bundles. 
This gives two different geometric  ways 
of computing the characters $g_{1,B}(\{t_i\}; CY)$ 
for general CY manifold: by Laurent expansion of the generating function for
homogeneous coordinates or by using the index theorem.

\subsubsection{Relation to Dimers and Integer Lattices}\label{pyramid}

It is instructive to draw the lattice of charges that the generating functions for fixed baryon numbers represent. Taking a power expansion of the form

\begin{equation}
g_{1,B}(t,x,y; {\cal C}) = \sum_{R, m_1, m_2} d_{R, B, m_1, m_2}t^{2R}x^{2m_1}y^{2m_2} ,
\end{equation}
we find that the coefficients $d_{R, B, m_1, m_2}$ are either $0$ or $1$. As such they form some kind of a fermionic condition on an occupation of the lattice point given by the three integer coordinates $2R, 2m_1, 2m_2$ (recall that $R, m_1, m_2$ admit half integral values and therefore twice their value is an integer). This phenomenon of fermionic exclusion was already observed for mesonic BPS gauge invariant operators in \cite{Benvenuti:2006qr} and persists for the case of baryonic BPS operators  \cite{Butti:2006au}. It is expected to persist for any singular toric non-compact CY manifold. In contrast, for the non-toric case, although there exist a lattice structure related to the generating function, the fermionic conditions seems to be lost \cite{Butti:2006nk}. Furthermore this lattice of points forms an ``integral conical pyramid" which is given by the intersection of two objects: A) a conical pyramid with a rectangular base, and B) the three dimensional integral lattice. Let us see this in some detail for several cases. Let us set $q=t^2$. With this definition the power of $q$ gives the $R$ charge of the corresponding operator. We first look at $B=0$ and expand equation (\ref{chebymesonid}) up to order $q^4$. We then write down a matrix which represents the two dimensional lattice in the $x,y$ coordinates.

\begin{equation}
\left(
\begin{array}{lllllllll}
 q^4 & 0 & q^4 & 0 & q^4 & 0 & q^4 & 0 & q^4 \\
 0 & q^3 & 0 & q^3 & 0 & q^3 & 0 & q^3 & 0 \\
 q^4 & 0 & q^4+q^2 & 0 & q^4+q^2 & 0 & q^4+q^2 & 0 & q^4
   \\
 0 & q^3 & 0 & q^3+q & 0 & q^3+q & 0 & q^3 & 0 \\
 q^4 & 0 & q^4+q^2 & 0 & q^4+q^2+1 & 0 & q^4+q^2 & 0 & q^4
   \\
 0 & q^3 & 0 & q^3+q & 0 & q^3+q & 0 & q^3 & 0 \\
 q^4 & 0 & q^4+q^2 & 0 & q^4+q^2 & 0 & q^4+q^2 & 0 & q^4
   \\
 0 & q^3 & 0 & q^3 & 0 & q^3 & 0 & q^3 & 0 \\
 q^4 & 0 & q^4 & 0 & q^4 & 0 & q^4 & 0 & q^4
\end{array}
\right) .
\end{equation}

\begin{figure}[h!!!]
\begin{center}
\includegraphics[scale=0.45]{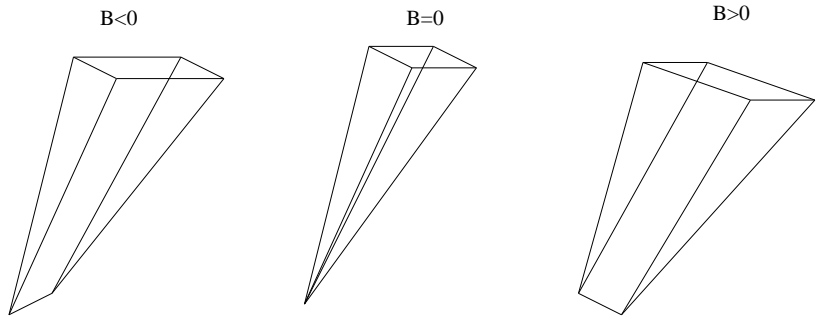} 
\caption{Examples of the ``integral conical pyramid"s for the conifold in the cases $B<0$, $B=0$, $B>0$.}\label{bariflop}
\end{center}
\end{figure}

The point in the middle contains $1=q^0$ and represents the identity operator. The 4 points around it contain a single power of $q$ and represent the 4 operators, $A_i B_j, i=1,2, j=1,2$, corresponding to the spin $(\frac{1}{2},\frac{1}{2})$ representation of $SU(2)_1\times SU(2)_2$. These points form an integral square of size 1. Next there are 9 points with $q^2$ that form a bigger integral square of size 2. These 9 points represent operators of the form $ABAB$ where we omit the indices but due to F term constraints we have 9 instead of 16 and transform now in the spin $(1,1)$ representation. Next 16 points in $q^3$ represent the operators $(AB)^3$ forming an integral square of size 3, etc. The generic case of $q^n$ will have $n^2$ points, representing the operators $(AB)^n$ which transform in the spin
$(\frac{n}{2},\frac{n}{2})$ representation, forming an integral square of size $n$. In fact this lattice is the weight lattice of the $SU(2)_1\times SU(2)_2$ irreducible representations of equal spin. We thus see that the lattice points are in correspondence with chiral operators in the $N=1, B=0$ theory and form a three dimensional lattice in the $q,x,y$ plane. Note that there are 4 diagonal lines which form the boundaries of this three dimensional lattice. These diagonals coincide with the $(p,q)$ web of the conifold theory, at the point where the size of the two cycle shrinks to zero, as in the middle of Figure \ref{bariflop}. We will soon identify the size of the two-cycle with the baryon number, or more precisely its absolute value. It is curious to note that this size is quantized, consistent with observations in \cite{Gopakumar:1998ki, Iqbal:2003ds}.

Let us turn to the case of baryonic charge $B=1$. We expand equation (\ref{chebyid}) to order $q^4$ and write the matrix for the two dimensional lattice in $x,y$,

\begin{equation}
q^{\frac{1}{2}}
\left(
\begin{array}{lllllllllll}
 q^4 & 0 & q^4 & 0 & q^4 & 0 & q^4 & 0 & q^4 & 0 & q^4 \\
 0 & q^3 & 0 & q^3 & 0 & q^3 & 0 & q^3 & 0 & q^3 & 0 \\
 q^4 & 0 & q^4+q^2 & 0 & q^4+q^2 & 0 & q^4+q^2 & 0 &
   q^4+q^2 & 0 & q^4 \\
 0 & q^3 & 0 & q^3+q & 0 & q^3+q & 0 & q^3+q & 0 & q^3 & 0
   \\
 q^4 & 0 & q^4+q^2 & 0 & q^4+q^2+1 & 0 & q^4+q^2+1 & 0 &
   q^4+q^2 & 0 & q^4 \\
 0 & q^3 & 0 & q^3+q & 0 & q^3+q & 0 & q^3+q & 0 & q^3 & 0
   \\
 q^4 & 0 & q^4+q^2 & 0 & q^4+q^2 & 0 & q^4+q^2 & 0 &
   q^4+q^2 & 0 & q^4 \\
 0 & q^3 & 0 & q^3 & 0 & q^3 & 0 & q^3 & 0 & q^3 & 0 \\
 q^4 & 0 & q^4 & 0 & q^4 & 0 & q^4 & 0 & q^4 & 0 & q^4
\end{array}
\right) .
\end{equation}

Note the factor of $q^{\frac{1}{2}}$ in front of this expression, indicating that there is one more $A$ field, carrying an $R$ charge $\frac{1}{2}$. For this case there are two lowest order lattice points with
$q^{\frac{1}{2}}$, corresponding to the operators $A_1, A_2$. These two points transform in the spin $(\frac{1}{2},0)$ representation and form an integral rectangle of sizes $2\times1$. At order $q^{\frac{3}{2}}$ we find 6 lattice points transforming in the spin $(1,\frac{1}{2})$ representation and forming an integral  rectangle of sizes $3\times2$, etc. At order $q^{n+\frac{1}{2}}$ we find $(n+2)(n+1)$ lattice points transforming in the spin $(\frac{n+1}{2},\frac{n}{2})$ and forming an integral rectangle of sizes $(n+2)\times(n+1)$. In summary the lattice we get is the weight lattice of all $SU(2)_1\times SU(2)_2$ representations with a difference of spin $\frac{1}{2}$. The 4 diagonal lines which form the boundaries of this three dimensional lattice now coincide with the $(p,q)$ web of the conifold theory, at a point where the size of the two cycle is non-zero, as in the right of Figure \ref{bariflop}.

As a last example let us turn to the case of baryonic charge $B=-2$. We expand equation (\ref{chebyid}) to order $q^4$ and write the matrix for the two dimensional lattice in $x,y$,

\begin{equation}
\left(
\begin{array}{lllllll}
 q^4 & 0 & q^4 & 0 & q^4 & 0 & q^4 \\
 0 & q^3 & 0 & q^3 & 0 & q^3 & 0 \\
 q^4 & 0 & q^4+q^2 & 0 & q^4+q^2 & 0 & q^4 \\
 0 & q^3 & 0 & q^3+q & 0 & q^3 & 0 \\
 q^4 & 0 & q^4+q^2 & 0 & q^4+q^2 & 0 & q^4 \\
 0 & q^3 & 0 & q^3+q & 0 & q^3 & 0 \\
 q^4 & 0 & q^4+q^2 & 0 & q^4+q^2 & 0 & q^4 \\
 0 & q^3 & 0 & q^3+q & 0 & q^3 & 0 \\
 q^4 & 0 & q^4+q^2 & 0 & q^4+q^2 & 0 & q^4 \\
 0 & q^3 & 0 & q^3 & 0 & q^3 & 0 \\
 q^4 & 0 & q^4 & 0 & q^4 & 0 & q^4
\end{array}
\right) .
\end{equation}

In this case there are two more $B$ fields than $A$ fields and hence there are three lowest order lattice points with $q^{{1}}$, corresponding to the operators $B_1B_1, B_1B_2, B_2B_2$. These three points transform in the spin $(0,1)$ representation and form an integral rectangle of sizes $1\times3$. At order $q^{{2}}$ we find 8 lattice points transforming in the spin $(\frac{1}{2},\frac{3}{2})$ representation and forming an integral  rectangle of sizes $2\times4$, etc. At order $q^{n+1}$ we find $(n+1)(n+3)$ lattice points transforming in the spin $(\frac{n}{2},\frac{n}{2}+1)$ and forming an integral rectangle of sizes $(n+1)\times(n+3)$. In summary the lattice we get is the weight lattice of all $SU(2)_1\times SU(2)_2$ representations with a negative difference of spin $1$. The 4 diagonal lines which form the boundaries of this three dimensional lattice now coincide with the $(p,q)$ web of the conifold theory after a flop transition, at a point where the size of the two cycle is non-zero, as in the left of Figure \ref{bariflop}.


We summarize this discussion by the observation that the boundaries of the conical pyramid are identified with the $(p,q)$ web of the conifold. This $(p,q)$ web can be resolved in two phases, separated by a flop transition. Each of these phases is characterized by the sign of the baryonic charge $B$ and the flop transition appears at $B=0$. In this sense one can identify the baryonic charge with the K\"ahler modulus for the conifold.

The relation to dimers then becomes natural. Let us fix a baryonic charge $B$. Take the tiling of the conifold, fix a reference tile and consider all open paths that begin with this tile, consists of $n_1$ arrows of type $A$ (horizontal) and $n_2$ arrows of type $B$ (vertical) such that the numbers $n_1$ and $n_2$ satisfy the condition $n_1-n_2=B$. For a given endpoint there may be more than one path which connects it to the reference tile. All such paths are equal by imposing the F-term conditions. See \cite{Hanany:2006nm} for a detailed explanation and \cite{Butti:2006hc,Butti:2006au} for the applications to the mesonic and baryonic BPS operators. The collection of endpoints then forms the two dimensional projection of the three dimensional ``conical pyramid" lattice discussed above.

It is instructive to relate this lattice to the Chebyshev polynomials discussed above. Let us denote the character $\chi_j (x) $ of the representation of spin $j$ by

\begin{equation}
\chi_j(x) = \sum_{m=-j}^j x^{2m} = \sum_{n=-\infty}^\infty d_n x^{n} .
\label{su2char}
\end{equation}

This character represents a one dimensional integer lattice of points running between $-2j$ and $2j$. We can call it the ``Integer Theta Function" (ITF) of the interval $I_j \equiv [-2j, 2j]$ since the coefficient $d_n$ is equal to 1 if $n\in I_j$ and 0 otherwise.
Equation (\ref{su2char}) is easily summed to give

\begin{equation}
\chi_j(x) = \frac{x^{2j+1} - x^{-2j-1} } {x - x^{-1} } = \frac {\sin [(2j+1)\theta]}{\sin\theta} = U_{2j}(\cos\theta),
\end{equation}
with $x=e^{i\theta}$. The last equality is precisely the definition of the Chebyshev polynomial of the second kind. We learn that the character of the $SU(2)$ representation of spin $j$ is the Chebyshev polynomial of the second kind, $U_{2j}(\cos\theta)$. Furthermore this polynomial gives the integer theta function (ITF) for the interval $I_j$.
Naturally, if we have a product of two characters $\chi_{j_1}(x) \chi_{j_2}(y)$ it will form the ITF on the two dimensional interval given by $I_{j_1,j_2} = I_{j_1}\times I_{j_2}$. Namely, the expansion

\begin{equation}
\chi_{j_1}(x) \chi_{j_2}(y) = \sum_{n_1=-\infty}^\infty \sum_{n_2=-\infty}^\infty d_{n_1, n_2} x^{n_1} y^{n_2},
\end{equation}
is such that $d_{n_1, n_2}=1$ if the integer point $(n_1,n_2)\in I_{j_1,j_2}$ and $d_{n_1, n_2}=0$ otherwise. At this point we observe that the building blocks of the generating functions for $N=1$ and fixed baryonic charge $g_{1,B}(t,x,y; {\cal C})$ in equations (\ref{chebyid}) are precisely the ITF of the two dimensional interval $I_{\frac {n+B}{2},\frac{n}{2}}$ for $B\ge0$ and $I_{\frac {n}{2},\frac{n+|B|}{2}}$ for $B\le0$. As we vary $n$ the size of the two dimensional interval varies such that the difference between the number of points on the two sides of the interval is equal to $B$. The collection of all such points forms the three dimensional integer lattice discussed above. In fact, using the new terminology, we can say that $g_{1,B}(t,x,y; {\cal C})$ is the ITF of the conical pyramid. This fact can be seen as the dimer realization of the geometric localization techniques \cite{Butti:2006au}.

\subsection{General Number of Branes, $N$, for the Conifold}

In the previous section we saw the generating function for the case of $N=1$ for the conifold. In this section we will develop the general $N$ case.
This is done using the Plethystic Exponential. Recall that in the case of baryon number $B=0$, namely for mesonic generating functions, \cite{Benvenuti:2006qr} the knowledge of the generating function for $N=1$ is enough to compute the generating function for any $N$. This is essentially due to the fact that the operators for finite $N$ are symmetric functions of the operators for $N=1$ and this is precisely the role which is played by the plethystic exponential -- to take a generating function for a set of operators and count all possible symmetric functions of it.

For the present case, in which the baryon number is non-zero, it turns out that the procedure is not too different than the mesonic case. One needs to have the knowledge of a single generating function, $g_{1,B}$, for one D brane, $N=1$ and for a fixed baryon number $B$, and this information is enough to compute all generating functions for any number of D branes and for a fixed baryonic number \cite{Butti:2006au}. We can summarize this information by writing an expression for the generating function of any number of branes and any baryonic number,

\begin{eqnarray}
\nonumber
g(\nu; t,b,x,y; {\cal C})
&=& \sum_{N=0}^\infty \nu^N g_N(t, b, x, y; {\cal C}) \\
\label{grandcancon}
&=& \sum_{B=-\infty}^\infty b^B \exp\biggl(\sum_{k=1}^\infty \frac{\nu^k}{k}g_{1,B}(t^k, x^k, y^k; {\cal C}) \biggr) \\
&=& \sum_{B=-\infty}^\infty b^B \exp\biggl(\sum_{k=1}^\infty \frac{\nu^k}{k}\frac{1}{2\pi i} \oint \frac {db'} {b'^{B+1}} g_1(t^k,b'^k,x^k,y^k; {\cal C}) \biggr) ,
\nonumber
\label{PEcon}
\end{eqnarray}
where the first equality indicates that this is a generating function for fixed number of branes, the second equality indicates that we are summing over all contributions of fixed baryonic numbers and that each contribution is the plethystic exponential of the generating function for one D brane and for fixed baryonic charge. The third equality expresses the generating function in terms of a multi-contour integral of the generating function for one D brane.
For completeness, using Equation (\ref{g1conitbxy}) and Equation (\ref{rescon}) we write down the generating function as explicit as possible,

\begin{eqnarray}
& &g(\nu; t,b,x,y; {\cal C}) = \\
\nonumber
\label{grandcanconex}
&=& \sum_{B=-\infty}^\infty b^B \exp\biggl(\sum_{k=1}^\infty \frac{\nu^k}{k}\frac{1}{2\pi i} \oint \frac {db'} {b'^{B+1}} \frac{1}{(1-t^k b'^k x^k) (1-\frac{t^k b'^k}{x^k})(1- \frac {t^k y^k}{b'^k}) (1-\frac{t^k}{b'^k y^k})}  \biggr) \\
\nonumber
&=& \sum_{B=0}^\infty b^B \exp\biggl(\sum_{k=1}^\infty
\frac{\nu^k t^{kB} x^{kB} } {k (1 - \frac{1}{x^{2k}}) (1-t^{2k} x^k y^k)  (1-\frac{t^{2k} x^k}{y^k}) }+ \frac{\nu^k t^{kB} x^{-kB}} { k (1 - x^{2k}) (1-\frac{t^{2k} y^k} {x^k})  (1-\frac{t^{2k}}{x^k y^k}) } \biggr) \\ \nonumber 
&+&\sum_{B=1}^{\infty} b^{-B} \exp \biggl ( \sum_{k=1}^\infty \frac{\nu^k t^{k B} y^{ k B} } { k (1-\frac{1}{y^{2k}}) (1-t^{2k} x^k y^k)  (1-\frac{t^{2k} y^k}{x^k}) }+ \frac{ \nu^k t^{k B} y^{- k B}} { k (1 - y^{2 k}) (1-\frac{t^{2 k} x^k} {y^k})  (1-\frac{t^{2 k}}{y^k x^k}) } \biggr) .
\nonumber
\end{eqnarray}

If we further ignore the $SU(2)$ quantum numbers and set $x=y=1$ we get

\begin{eqnarray}
g(\nu; t,b,1,1; {\cal C})
\label{grandcancon11}
&=& \exp\biggl(\sum_{k=1}^\infty
\frac{ \nu^k (1+t^{2k})} { k (1- t^{2k})^3} \biggr) \\ &+& \sum_{B=1}^\infty ( b^B + b^{-B} )\exp\biggl(\sum_{k=1}^\infty
\frac{ \nu^k t^{ k B } (1+t^{2k})} { k (1- t^{2k})^3}+ \frac{ B \nu^k t^{ k B } } { k (1- t^{2 k} )^2 }\biggr) .
\nonumber
\end{eqnarray}

The exponential terms have an equivalent representation as infinite products and take the form

\begin{eqnarray}
g(\nu; t,b,1,1; {\cal C})
= \prod_{n=0}^\infty \frac{1}{(1-\nu t^{2n})^{(n+1)^2}}
+ \sum_{B=1}^\infty ( b^B + b^{-B} ) \prod_{n=0}^\infty \frac {1} {(1-\nu t^{2n+B})^{(n+1)(n+B+1)} } .
\nonumber
\end{eqnarray}

\subsubsection{$N=2$ for the Conifold}
\label{Nis2}

Taking the expansion of Equation (\ref{PEcon}) to second order in the chemical potential $\nu$ for the number of D branes, we find

\begin{eqnarray}
g_2(t, b, x, y; {\cal C}) &=& \sum_{B=-\infty}^\infty b^B \frac{1}{2}\biggl[\frac{1}{2\pi i} \oint \frac {db'} {b'^{B+1}} g_1(t,b',x,y; {\cal C}) \biggr]^2 \\ \nonumber
&+& \sum_{B=-\infty}^\infty b^B \frac{1}{2}\frac{1}{2\pi i} \oint \frac {db'} {b'^{B+1}} g_1(t^2,b'^2,x^2,y^2; {\cal C}) = F_1 + F_2 ,
\end{eqnarray}
and we divided the contribution into two terms, $F_1$ and $F_2$. We next use the identity

\begin{eqnarray}
\sum_{B=-\infty}^\infty b^B \frac{1}{2\pi i} \oint \frac {db'} {b'^{B+1}} = \delta(b-b').
\end{eqnarray}
and evaluate $F_2$ to be

\begin{eqnarray}
F_2=\frac{1}{2}g_1(t^2,b^2,x^2,y^2; {\cal C}) .
\end{eqnarray}

$F_1$ is slightly more involved since it contains two contour integrals but still is relatively easy to evaluate,

\begin{eqnarray}
\nonumber
F_1 &=& \sum_{B=-\infty}^\infty b^B \frac{1}{2}\frac{1}{2\pi i} \oint \frac {db'} {b'^{B+1}} g_1(t,b',x,y; {\cal C}) \frac{1}{2\pi i} \oint \frac {db''} {b''^{B+1}} g_1(t,b'',x,y; {\cal C}) \\ \nonumber
&=& \sum_{B=-\infty}^\infty b^B \frac{1}{2}\frac{1}{(2\pi i)^2} \oint \frac {db'} {b'^{B+1}} \oint \frac {db''} {b''^{B+1}} g_1(t,b',x,y; {\cal C})  g_1(t,b'',x,y; {\cal C}) \\ \nonumber
&=& \sum_{B=-\infty}^\infty b^B \frac{1}{2}\frac{1}{(2\pi i)^2} \oint \oint \frac {ds'} {s'^{B+1}} \frac {ds} {s} g_1(t,s s',x,y; {\cal C})  g_1(t,\frac{s'}{s},x,y; {\cal C}) \\ \nonumber
&=& \frac{1}{2}\frac{1}{(2\pi i)} \oint \frac {ds} {s} g_1(t,b s,x,y; {\cal C})  g_1(t,\frac{b}{s},x,y; {\cal C}) . \\ \nonumber
\end{eqnarray}

We can now collect all the terms and get an expression for $N=2$,

\begin{equation}
g_2(t, b, x, y; {\cal C}) = \frac{1}{2}g_1(t^2,b^2,x^2,y^2; {\cal C})
+ \frac{1}{2}\frac{1}{(2\pi i)} \oint \frac {ds} {s} g_1(t,b s,x,y; {\cal C})  g_1(t,\frac{b}{s},x,y; {\cal C}) .
\label{g2coni}
\end{equation}

We can recall the chemical potentials for counting the $A$ and $B$ fields and rewrite

\begin{equation}
g_2(t_1, t_2, x, y; {\cal C}) = \frac{1}{2}g_1(t_1^2, t_2^2, x^2, y^2; {\cal C})
+ \frac{1}{2}\frac{1}{(2\pi i)} \oint \frac {ds} {s} g_1(t_1 s,\frac{t_2}{s}, x, y; {\cal C})
g_1(\frac{t_1}{s},t_2 s, x, y; {\cal C}) .
\end{equation}

Let us evaluate the second term and write it explicitly using equation (\ref{g1coni}).

\begin{equation}
\frac{1}{2}\frac{1}{(2\pi i)} \oint \frac {ds} {s}
\frac{1}{(1-t_1s x) (1-\frac{t_1s}{x})(1-\frac{t_2 y}{s}) (1-\frac{t_2}{s y})}
\frac{1}{(1-\frac{t_1 x}{s}) (1-\frac{t_1}{s x})(1-t_2 s y) (1-\frac{t_2 s}{y})} .
\end{equation}

The residue integral now gets contributions from 4 different points at

\begin{equation}
s=t_1 x, \qquad s=\frac{t_1}{x}, \qquad s=t_2 y, \qquad s=\frac{t_2}{y}.
\end{equation}

The computations are a bit lengthy but after some work we get an expression for the generating function for BPS operators on N=2 D branes probing the conifold

$$
g_2(t_1, t_2, x, y; {\cal C}) =\,\,\qquad\qquad\qquad\qquad\qquad\qquad\qquad\qquad\qquad\qquad\qquad\qquad\qquad\qquad$$
$$
\frac
{1-t_1^3t_2^3(1-t_1^2)(1-t_2)^2\chi_{\frac{1}{2}}(x)\chi_{\frac{1}{2}}(y)-t_1^2t_2^2[(1-t_1^2)t_2^2\chi_1(x)+(1-t_2^2)t_1^2\chi_1(y)]+t_1^4t_2^4(1-t_1^2-t_2^2)}{(1-t_1^2)(1-t_2^2)( 1 - t_1^2 x^2 )( 1 - t_2^2 y^2 )(1-\frac{t_1^2}{x^2})(1-\frac{t_2^2}{y^2})(1-t_1t_2xy)(1-\frac{t_1t_2x}{y})(1-\frac{t_1t_2y}{x})(1-\frac{t_1t_2}{xy})} . \nonumber
$$


We can ignore the $SU(2)$ weights by setting $x=y=1$ and by using $\chi_j(1) = 2j+1$,

\begin{equation}\label{N2con}
g_2(t_1, t_2, 1, 1; {\cal C}) = \frac{1+t_1t_2+t_1^2t_2^2[1-3(t_1^2+t_2^2)]+t_1^3t_2^3(t_1^2+t_2^2-3)+4t_1^4t_2^4}
{(1-t_1^2)^3(1-t_1t_2)^3(1-t_2^2)^3} .
\end{equation}

Similar expressions can be obtained for $N=3$ and greater values of $N$. The
properties of these generating functions for different values of $N$ are 
discussed in Section \ref{PLlog}.

\subsection{The Plethystic Exponential, the Baryonic Chiral Ring and the Geometric Quantization Procedure}

It should be clear from the previous sections that the knowledge of the generating function $g_{1,B}(q;\mathcal{C})$ for $N=1$ and fixed baryonic number $B$ is enough to compute the generating function for any $N$ and $B$. Intuitively this is essentially due to the fact that the chiral BPS operators for finite $N$ are symmetric functions of the operators for $N=1$. The plethystic exponential has the role of taking a generating function for a set of operators and of counting all possible symmetric functions of them, therefore it allows to pass from $g_{1,B}(q;\mathcal{C})$ to $g_{N, N B}(q;\mathcal{C})$. In this section we want to explain in detail how this procedure works for the simple case of the conifold for baryonic charge  $N$, and the geometric realization of this procedure.

Using the elementary  fields $(A_1,B_1,A_2,B_2)$ transforming under the gauge group $SU(N) \times SU(N)$ and with baryonic charge $(1,-1,1,-1)$ one can build the basic baryonic operators with baryonic charge $N$:

\begin{equation}
\label{exbar}
\epsilon^1 _{p_1,...,p_N}  \epsilon_2 ^{k_1,...,k_N} (A_{i_1})^{p_1}_{k_1}... (A_{i_N})^{p_N}_{k_N} = (\det A)_{ ( i_1,...,i_N ) } . \nonumber \\
\end{equation}

Due to the $\epsilon$ antisymmetrization, these operators are symmetric in the exchange of the $A_i$, and transform under the spin $(\frac{N}{2},0)$ representation of $SU(2)_1 \times SU(2)_2$.    
It is clearly possible to construct other operators with $B=N$ with bigger dimension. Defining the operators \cite{Berenstein:2002ke,Beasley:2002xv}

\begin{equation}\label{AAA}
A_{I;J}= A_{i_1}B_{j_1}...A_{i_m}B_{j_m}A_{i_{m+1}}
\end{equation}
the generic baryonic operator with baryonic charge $N$ is indeed:

\begin{equation}
\label{genbarA}
\epsilon^1 _{p_1,...,p_N}  \epsilon_2 ^{k_1,...,k_N} (A_{I_{1};J_{1}})^{p_1}_{k_1}... (A_{I_{N};J_{N}})^{p_N}_{k_N}\, . 
\end{equation}

Thanks to the $\epsilon$ symbols these operators are completely symmetric in the exchange of the $A_{I;J}$'s. Now we want to understand the role of the plethystic exponential in the problem of counting all the operators of the form (\ref{genbarA}). 
It is clear that the generating function counting all the possible BPS operators with baryonic charge $N$ is the one that counts all the possible symmetric products of $N$ elements of the form (\ref{AAA}). The BPS operators (\ref{AAA}) are in correspondence with points inside an integral conical pyramid (see Section \ref{pyramid}), that we denote $P_{B=1}$. Therefore counting the operators (\ref{AAA}) is the same as counting the integer points $m=(m_1,m_2,m_3) \in \mathbb{Z}^3$ inside $P_{B=1}$, and the counting procedure for generic $N$ is simply achieved by:
\begin{equation}\label{PLnu}
\prod_{m \in P_{B=1}} \frac{1}{1-\nu q^m}= 
\sum_{N = 0}^{\infty}\nu^N \hbox{ (all symmetric products of $N$ elements in $P_{B=1}$) } ,
\end{equation}
where we have introduced the chemical potential $\nu$ for the number of branes, and $q=(q_1,q_2,q_3)$ are the chemical potentials for the $T^3$ toric action.
From this expansion it is clear that the LHS of Equation (\ref{PLnu}) is the generating function for all the possible symmetric products of $N$ elements inside $g_{1,1}(q;\mathcal{C})$. 
The RHS of Equation (\ref{PLnu}) is what we write as $\sum_{N=0}^{\infty}\nu^N g_{N,N}(q;\mathcal{C})$, and it is easy to show that:

\begin{equation}\label{PLnufr}
\prod_{m \in P_{B=1}} \frac{1}{1-\nu q^m}= \exp{\Big(\sum_{k=1}^{\infty} \frac{\nu ^{k}}{k} g_{1,1}(q^k;\mathcal{C})}\Big)
\end{equation}

The RHS of Equation (\ref{PLnufr}) is the definition of the plethystic exponential, hence we have the relation:

\begin{equation}\label{PLnuex}
\hbox{PE$_{\nu}$} [ g_{1,1}(q;\mathcal{C})]= \sum_{N=0}^{\infty}\nu^N g_{N,N}(q;\mathcal{C}) .
\end{equation}

This is the physical justification of the use of the plethystic exponential in the baryonic counting problem: the generic BPS operator of the gauge theory in the presence of $N$ D3 branes, Equation (\ref{genbarA}), is a symmetric function of the chiral operators in the case of $N=1$, Equation (\ref{AAA}). Hence once we know $g_{1,B}(q;\mathcal{C})$ we obtain $g_{N,NB}(q;\mathcal{C})$ by counting all possible symmetric functions of the operators in $g_{1,B}(q;\mathcal{C})$, and this is exactly the role of the plethystic exponential.

It is also possible to relate the plethystic exponential to the result of the geometric quantization procedure of a system of D3 branes wrapped on three cycles inside $T^{1,1}$. It is known that the gauge theory we have discussed so far is dual to Type IIB string theory on the $AdS_5 \times T^{1,1}$ supergravity background. In the geometric side the baryonic operators correspond to D3 branes wrapped over non trivial three cycles in $T^{1,1}$. Let us briefly discuss how this correspondence works in the case of baryonic number $N$. The supersymmetric D3 brane wrapping three-cycles in $T^{1,1}$ are in one to one correspondence with the holomorphic surfaces $S$ in the real cone $C(T^{1,1})$ over $T^{1,1}$ \cite{Mikhailov:2000ya,Beasley:2002xv}. As explained in Section (\ref{geometricamusement}), we can associate to every vertex of the toric diagram a global homogeneous coordinate $x_i$. In the conifold case the homogeneous coordinates $(x_1,x_2,x_3,x_4)$ with charges $(1,-1,1,-1)$ can be put in one-to-one correspondence with the elementary fields $(A_1,B_1,A_2,B_2)$  which have indeed the same charges. Hence all the supersymmetric configurations of D3 branes wrapped in $T^{1,1}$ with $B=1$ are zero loci of homogeneous polynomials of degree one:  
\begin{eqnarray}
P_{B=1}(x_1,x_2,x_3,x_4) &\equiv&  h_1 x_1 + h_3 x_3 + \nonumber \\ 
                              & &  h_{11;2} x_1^2 x_2 + h_{13;2} x_1 x_3 x_2 + h_{33;2} x_3^2 x_2 +  \nonumber \\ 
                              & &  h_{11;4} x_1^2 x_4 + h_{13;4} x_1 x_3 x_4 + h_{33;4} x_3^2 x_4 + ...   
\end{eqnarray}
where $h_{m;n}$'s are complex numbers and parametrize the supersymmetric D3 brane embeddings.
It turns out that the phase space of this classical system of D3 branes is $\mathbb{CP}^{\infty}$ and the $h_{m;n}$'s are its homogeneous coordinate \cite{Beasley:2002xv}. It is possible to quantize the classical phase space using the geometric quantization techniques \cite{Wood,Beasley:2002xv}. The result is that the BPS Hilbert space corresponding to the classical phase space is spanned by the degree $N$ polynomials in the homogeneous coordinates $h_{m;n}$, where $N$ is the value of the integral of the five form $F_5$ over $T^{1,1}$, and it correspond to the number of colors in the dual gauge theory. These polynomials are clearly completely symmetric in the exchange of the $h_{m;n}$'s and one can write them as the symmetric states: 
\begin{equation}
| h_{m_1;n_1}, h_{m_2;n_2},...,h_{m_N;n_N} \rangle 
\label{bpsstate}
\end{equation} 
In this geometric language every $h_{m;n}$ correspond to a section of the line bundle $\mathcal{O}(B=1)$ and the problem of counting all these sections is explained in detail in \cite{Butti:2006au} and reviewed in Section \ref{geometricamusement}. Let us denote the generating function for this counting problem $Z_{1,1}(q;\mathcal{C})$. The geometric quantization procedure prescribes that (\ref{bpsstate}) are a basis of the Hilbert space for the BPS wrapped D3 branes, and hence of all the possible symmetric products of $N$ of the sections of $\mathcal{O}(B=1)$. As explained above, it is by now clear that the role of the plethystic exponential is to pass from the generating function of the global sections of $\mathcal{O}(B=1)$ to the generating function of their $N$-times symmetric products:
\begin{equation}
\hbox{PE$_{\nu}$}[Z_{1,1}(q;\mathcal{C})]= \sum_{N=0}^{\infty} \nu^N Z_{N,N}(q;\mathcal{C})
\end{equation}
Hence in the geometric picture the appearance of the plethystic exponential is the realization of the geometric quantization procedure over the phase space of a system of wrapped D3 branes in the internal geometry.

Clearly, there exists a precise relation between the two counting problems\cite{Beasley:2002xv}. From the identification of the elementary fields of the conifold with the homogeneous coordinates $x_i$, one has the obvious association between the homogeneous coordinates $h_{m;n}$ over the phase space and the chiral operators with $B=1$:

\begin{equation} 
h_{i_1,...,i_{m+1};j_1,...,j_m} \leftrightarrow A_{i_1}B_{j_1}...A_{i_m}B_{j_m}A_{i_{m+1}}
\end{equation}
and hence the correspondence between the BPS states in $AdS_5\times T^{1,1}$ and the baryonic operators in the dual field theory \cite{Beasley:2002xv}:  
\begin{equation}
| h_{I_1;J_1},...,h_{I_N;J_N} \rangle  \leftrightarrow \epsilon^1\epsilon_2 (A_{I_{1};J_{1}},...,A_{I_{N};J_{N}}) 
\end{equation}
Therefore the interpretation of the plethystic exponential in these counting problems is two-fold: in the gauge theory side it is the realization that the generic BPS baryonic operator is a symmetric product of some basic building blocks; in the geometric side it is the direct outcome of the geometric quantization procedure of a system of D3 branes. The two approaches are related by the $AdS/CFT$ correspondence\footnote{ In the generic toric case there exist a relation between the two counting problem, but it is more subtle, mainly due to the fact that the correspondence between the homogeneous coordinates $x_i$ and the elementary fields of the dual gauge theory is not one to one.}:
\begin{equation}\label{relgg}
g_{1,1}(q;\mathcal{C}) = Z_{1,1}(q;\mathcal{C})  \hbox{             }  \hbox{           }  \hbox{ , } \hbox{             } \hbox{             } g_{N,N}(q;\mathcal{C}) = Z_{N,N}(q;\mathcal{C})
\end{equation}

\subsection{The Structure of the Chiral Ring for the Conifold: the Plethystic Log}\label{PLlog}

Now that we have under control the generating functions for the conifold for generic baryonic number $B$ and number of branes $N$, it is interesting to understand the structure of the chiral ring for different values of $N$.

For simplicity we consider the generating functions for the conifold $g_N(t_1,t_2;\cal{C})$ with $x=1$, $y=1$. In this way it counts only the number of fields $A_i$ ($t_1$) and $B_i$ ($t_2$) in the gauge invariant operators without taking into account their weights under the global $SU(2)$ symmetries.
Our procedure will be to take the plethystic logarithm (\hbox{PE$^{-1}$}, or equivalently \hbox{PL}) of the generating function $g_N(t_1,t_2;\cal{C})$. This operation is the inverse function of the plethystic exponential and is defined in the following way:

\begin{equation}
\hbox{PE$^{-1}$}[g_N(t_1,t_2;{\cal C})] \equiv \sum_k^{\infty} \frac{\mu (k)}{k} \log (g_N(t_1^k,t_2^k;\cal{C})) ,
\end{equation}
where $\mu (k)$ is the M\"obius function\footnote{\begin{equation}
\mu(k) = \left\{\begin{array}{lcl}
0 & & k \mbox{ has one or more repeated prime factors}\\
1 & & k = 1\\
(-1)^n & & k \mbox{ is a product of $n$ distinct primes}
\end{array}\right. \ .
\end{equation}}.

The important fact about this operator is that acting with \hbox{PE$^{-1}$} on a generating function we obtain the generating series for the generators and the relations in the chiral ring.
The result is generically\footnote{In the $AdS/CFT$ correspondence the moduli space of the gauge theory 
is typically not a complete intersection variety} an infinite series in which the first terms with the plus sign give the basic generators while the first terms with the minus sign give the relations between these basic generators. Then there is an infinite series of terms with plus and minus signs due to the fact that the moduli space of vacua is not a complete intersection and the relations in the chiral ring are not trivially generated by the relations between the basic invariants, but receives stepwise corrections at higher degrees.

Let us start with the simplest case $N=1$. In this case the generating function for the conifold is simply:
\begin{equation}
g_1(t_1,t_2;{\cal C})= \frac{1}{(1 - t_1)^2(1 - t_2)^2} .
\end{equation} 
Taking the plethystic logarithm we obtain:

\begin{equation}
\hbox{PE$^{-1}$} [g_1(t_1,t_2;{\cal C})]= 2 t_1 + 2 t_2 .
\end{equation}

This means that in the case $N=1$ the chiral ring is freely generated by $A_1$, $A_2$ and $B_1$, $B_2$ without any relations. Indeed in this case the gauge theory is Abelian and we have no matrix relations, the superpotential is zero and we have no $F$-term relations.

We can now pass to the more interesting case of $N=2$. The generating function is given in Equation (\ref{N2con}):

\begin{equation}
g_2(t_1,t_2;{\cal C})= \frac{1 + t_1 t_2 + t_1^2 t_2^2 - 3 t_1^4 t_2^2 - 3 t_1^2 t_2^4 + t_1^5 t_2^3 + t_1^3 t_2^5  - 3 t_1^3 t_2^3 + 4 t_1^4 t_2^4}{(1 - t_1^2)^3(1 - t_1 t_2)^3 (1 - t_2^2)^3} .
\end{equation} 

The first terms of its  plethystic logarithm are:

\begin{equation}\label{PLg2}
\hbox{PE$^{-1}$}[g_2(t_1,t_2;{\cal C})] = 3t_1^2 + 4t_1 t_2 + 3t_2^2 - 3 t_1^4 t_2^2 - 4t_1^3 t_2^3 -3t_1^2 t_2^4 + ...
\end{equation} 

As explained above the interpretation of equation (\ref{PLg2}) is: the first three monomials are the basic generators of the chiral ring and we can recognize them in the following gauge invariant operators in the quiver theory
:

\begin{eqnarray}\label{basicgen}
3 t_1^2 & \rightarrow & \epsilon \epsilon A_1 A_1 \hbox{ , } \epsilon \epsilon A_1 A_2 \hbox{ , } \epsilon \epsilon A_2 A_2 \nonumber \\
3 t_2^2 & \rightarrow & \epsilon \epsilon B_1 B_1 \hbox{ , } \epsilon \epsilon B_1 B_2 \hbox{ , } \epsilon \epsilon B_2 B_2 \nonumber \\
4t_1 t_2 & \rightarrow & \hbox{Tr}(A_1 B_1) \hbox{ , } \hbox{Tr}(A_1 B_2) \hbox{ , } \hbox{Tr}(A_2 B_1) \hbox{ , } \hbox{Tr}(A_2 B_2) 
\end{eqnarray}
where the indices contraction between fields and epsilon symbols is implicit. These operators transform under spin (1,0), (0,1), and $(\frac{1}{2},\frac{1}{2})$ of $SU(2)_1\times SU(2)_2$, respectively.

The second three monomials give the quantum number of the relations between the basic generators of the chiral ring. The presence of these relations means that the chiral ring in the case $N=2$ is not freely generated. It is possible to understand this fact looking at the higher degree gauge invariant fields in the chiral ring. At order $t_1^4 t_2^2$ using the ten basic generators we can write the operators:
\begin{equation}\label{relb}
(\epsilon \epsilon A_i A_j)^2 (\epsilon \epsilon B_k B_l) \hbox{ , } (\hbox{Tr}(A_i B_j))^2 (\epsilon \epsilon A_k A_l) \hbox{ :          18 + 30 operators}\end{equation}
 Using the tensor relation
\begin{equation}\label{tensrel}
\epsilon_{\alpha_1...\alpha_N}\epsilon^{\beta_1...\beta_N} = \delta_{[ \alpha_1}^{\beta_1}...\delta_{\alpha_N]}^{\beta_N}
\end{equation}
and some tensor algebra we can rewrite the gauge invariants in equation (\ref{relb}) in terms of the operators:
\begin{equation}\label{relc}
\epsilon \epsilon (A_i B_j A_k)(A_l B_m A_n) \hbox{ , } \epsilon \epsilon (A_i B_j A_k B_l A_m)(A_n) \hbox{ :          21 + 24 operators}
\end{equation}
Hence it is possible to write the $48$ operators in equation (\ref{relb}) in terms of the $45$ operators in equation (\ref{relc}). This means that there exist 
at least $3$ relations with quantum numbers $t_1^4t_2^2$ between the ten basic generators in Equation (\ref{basicgen}). One can check that the relations are exactly three and these are precisely the ones predicted by the term $-3t_1^4 t_2^2$ in Equation (\ref{PLg2}).
To justify the term  $-3t_1^2 t_2^4$ in (\ref{PLg2}) one works in the same way as before exchanging the $A$ fields with the $B$ fields.
In a similar way we can justify the term $-4t_1^3 t_2^3$ in (\ref{PLg2}). Using the ten basic generators we can write the following $56$ operators at level $t_1^3 t_2^3$:
\begin{equation}\label{reld}
(\epsilon \epsilon A_i A_j)(\epsilon \epsilon B_k B_l)\hbox{Tr}(A_m B_n) \hbox{ , } (\hbox{Tr}(A_i B_j))^3 \hbox{ :          36 + 20 operators}
\end{equation}
these can be written in terms of the $52$ operators:
\begin{equation}\label{rele}
\epsilon \epsilon (A_i B_j A_k B_l)(A_m B_n) \hbox{ , } \epsilon \epsilon (A_i B_j A_k B_l A_m B_n)(1) \hbox{ :           36 + 16 operators}
\end{equation}
where $1$ is the identity operator.

Hence we see that in field theory there exist $4$ relations with the quantum numbers $t_1^3t_2^3$, and these are the ones predicted by the  plethystic logarithm in equation (\ref{PLg2}).

The higher monomials in equation (\ref{PLg2}) mean that the moduli space of the conifold at $N=2$ is not a complete intersection.

At $N=2$ one may expect the presence of the mesonic operators $\hbox{Tr}(ABAB)$ among the basic generators of the chiral ring, but it is easy to show that:

\begin{equation}\label{abab}
(\epsilon \epsilon A_i A_j )(\epsilon \epsilon B_k B_l ) = (\hbox{Tr}(A_iB_j))^2 - \hbox{Tr}(A_iB_kA_jB_l) ,
\end{equation}
and hence we conclude that $\hbox{Tr}(ABAB)$ do not appear independently, but are generated by the ten basic gauge invariant operators in (\ref{basicgen}) as predicted by the plethystic logarithm in Equation (\ref{PLg2}).

To check the consistency of the discussion about the basic generators and their relations one can expand the generating function $g_2(t_1,t_2;{\cal C})$ at the first few orders:
\begin{eqnarray}\label{g2ffew}
g_2(t_1,t_2;{\cal C})= 1 + 3t_1^2 + 4t_1 t_2+ 3t_2^2  + 6 t_1^4  + 12t_1^3t_2 + 19t_1^2t_2^2 +  12t_1 t_2^3 + 6t_2^4 + \nonumber\\ 10t_1^6+ 24t_1^5t_2 + 45t_1^4t_2^2  + 52 t_1^3t_2^3  + 45 t_1^2 t_2^4 + 24t_1t_2^5 + 10t_2^6 + ...
\end{eqnarray}
and try to construct the gauge invariant operators associated with each monomial.

It is convenient to form this expansion into a two dimensional lattice in the $t_1, t_2$ coordinates,

\begin{equation}
\left(
\begin{array}{lllllllll}
 1 & 0 & 3 & 0 & 6 & 0 & 10 & 0 & 15 \\
 0 & 4 & 0 & 12 & 0 & 24 & 0 & 40 & 0 \\
 3 & 0 & 19 & 0 & 45 & 0 & 81 & 0 & 127 \\
 0 & 12 & 0 & 52 & 0 & 112 & 0 & 192 & 0 \\
 6 & 0 & 45 & 0 & 134 & 0 & 258 & 0 & 417 \\
 0 & 24 & 0 & 112 & 0 & 280 & 0 & 504 & 0 \\
 10 & 0 & 81 & 0 & 258 & 0 & 554 & 0 & 934 \\
 0 & 40 & 0 & 192 & 0 & 504 & 0 & 984 & 0 \\
 15 & 0 & 127 & 0 & 417 & 0 & 934 & 0 & 1679
\end{array}
\right) .
\end{equation}

The $1$ is the identity operator, the quadratic monomials are the basic generators already discussed, at the quartic order we begin to have composite operators made out of the 10 generators. Let us count them taking into account their symmetry properties:
\begin{eqnarray}\label{quart}
6 t_1^4 & \rightarrow & (\epsilon \epsilon A_i A_j)^2  \hbox{ :         6 operators} \nonumber \\
12 t_1^3t_2 & \rightarrow & (\epsilon \epsilon A_i A_j) \hbox{Tr}(A_k B_l) \hbox{ :        12 operators} \nonumber \\
19 t_1^2 t_2^2 & \rightarrow & (\epsilon \epsilon A_i A_j)  (\epsilon \epsilon B_k B_l) \hbox{ , } (\hbox{Tr}(A_i B_j))^2\hbox{ :         9 + 10 operators} \nonumber\\
12 t_1t_2^3 & \rightarrow & (\epsilon \epsilon B_i B_j) \hbox{Tr}(A_k B_l) \hbox{ :         12 operators} \nonumber \\
6 t_2^4 & \rightarrow & (\epsilon \epsilon B_i B_j)^2 \hbox{ :         6 operators} 
\end{eqnarray}
Thus, the counting as encoded in the generating function and the 
explicit counting in the gauge theory nicely agree.
It is interesting to do the same analysis at dimension six, because at this level the relations among the basic generators enter into the game (see equation (\ref{PLg2})):
\begin{eqnarray}\label{sixth}
10t_1^6& \rightarrow & (\epsilon \epsilon A_i A_j)^3  \nonumber \\ & & \hbox{          10 operators} \nonumber \\
24t_1^5t_2 & \rightarrow & (\epsilon \epsilon A_i A_j)^2  \hbox{Tr}(A_k B_l) \nonumber \\ & & \hbox{         24 operators} \nonumber \\
45t_1^4 t_2^2 & \rightarrow & (\epsilon \epsilon A_i A_j)^2  (\epsilon \epsilon B_k B_l) \hbox{ , } (\hbox{Tr}(A_i B_j))^2(\epsilon \epsilon A_n A_m) \nonumber \\ & & \hbox{        18 + 30 operators - 3 relations = 45 operators } \nonumber\\
52 t_1^3t_2^3 & \rightarrow & (\epsilon \epsilon A_i A_j)  (\epsilon \epsilon B_k B_l) \hbox{Tr}(A_n B_m) \hbox{ , } (\hbox{Tr}(A_i B_j))^3  \nonumber \\ & & \hbox{        36 + 20 operators - 4 relations = 52 operators } \nonumber\\
45t_1^2 t_2^4 & \rightarrow & (\epsilon \epsilon B_i B_j)^2  (\epsilon \epsilon A_k A_l) \hbox{ , } (\hbox{Tr}(A_i B_j))^2 (\epsilon \epsilon B_n B_m)  \nonumber \\ & & \hbox{        18 + 30 operators - 3 relations = 45 operators } \nonumber\\
24t_1t_2^5 & \rightarrow & (\epsilon \epsilon B_i B_j)^2 \hbox{Tr}(A_k B_l)  \nonumber \\ & & \hbox{         24 operators} \nonumber \\
10t_2^6& \rightarrow & (\epsilon \epsilon B_i B_j)^3   \nonumber \\ & & \hbox{          10 operators} 
\end{eqnarray}
Also at this level the computation with the generating function and with the conformal field theory techniques, taking into account the relations between the basic generators previously explained, nicely agree.

It is now interesting to give a look at the case of $N=3$. This is the first level where non factorizable baryons appear \cite{Berenstein:2002ke,Beasley:2002xv}. 
These types of baryons are the ones that cannot be written as a product of mesonic operators (traces of fields) and the basic baryons (for example $ \epsilon \epsilon A_i A_j A_k$ at the level $N=3$ is a basic baryon).
The generating function for $N=3$ has the form:
\begin{equation}\label{g3con}
g_3(t_1,t_2;{\cal C})=\frac{F(t_1,t_2)}{(1-t_1^3)^4(1 - t_1 t_2)^3 (1-t_1^2 t_2^2)^3(1- t_2^3)^4} ,
\end{equation}
where $F(t_1,t_2)$ is a polynomial in $t_1$, $t_2$.
The first few terms in the $t_1, t_2$ lattice look like

\begin{equation}
\left(
\begin{array}{lllllllllll}
 1 & 0 & 0 & 4 & 0 & 0 & 10 & 0 & 0 & 20 & 0 \\
 0 & 4 & 0 & 0 & 18 & 0 & 0 & 48 & 0 & 0 & 100 \\
 0 & 0 & 19 & 0 & 0 & 78 & 0 & 0 & 198 & 0 & 0 \\
 4 & 0 & 0 & 72 & 0 & 0 & 260 & 0 & 0 & 624 & 0 \\
 0 & 18 & 0 & 0 & 224 & 0 & 0 & 738 & 0 & 0 & 1686 \\
 0 & 0 & 78 & 0 & 0 & 620 & 0 & 0 & 1854 & 0 & 0 \\
 10 & 0 & 0 & 260 & 0 & 0 & 1545 & 0 & 0 & 4246 & 0 \\
 0 & 48 & 0 & 0 & 738 & 0 & 0 & 3508 & 0 & 0 & 8958 \\
 0 & 0 & 198 & 0 & 0 & 1854 & 0 & 0 & 7414 & 0 & 0 \\
 20 & 0 & 0 & 624 & 0 & 0 & 4246 & 0 & 0 & 14728 & 0 \\
 0 & 100 & 0 & 0 & 1686 & 0 & 0 & 8958 & 0 & 0 & 27729
\end{array}
\right) .
\end{equation}

We now take the plethystic logarithm to understand the structure of basic generators at level $N=3$:

\begin{equation}\label{PLg3con}
\hbox{PE$^{-1}$} [g_3(t_1,t_2;{\cal C})] = 4t_1^3 + 2 t_1^4 t_2  + 2t_1t_2^4 + 4t_2^3 + 4 t_1 t_2  + 9t_1^2 t_2^2+...
\end{equation}

with the dots meaning the presence of relations among the basic generators at higher orders in $t_1$, $t_2$. On the $t_1,t_2$ lattice this expression looks like this

\begin{equation}
\left(
\begin{array}{lllllllllll}
 0 & 0 & 0 & 4 & 0 & 0 & 0 & 0 & 0 & 0 & 0 \\
 0 & 4 & 0 & 0 & 2 & 0 & 0 & 0 & 0 & 0 & 0 \\
 0 & 0 & 9 & 0 & 0 & -6 & 0 & 0 & -3 & 0 & 0 \\
 4 & 0 & 0 & 0 & 0 & 0 & -18 & 0 & 0 & 12 & 0 \\
 0 & 2 & 0 & 0 & -26 & 0 & 0 & 24 & 0 & 0 & 27 \\
 0 & 0 & -6 & 0 & 0 & -16 & 0 & 0 & 136 & 0 & 0 \\
 0 & 0 & 0 & -18 & 0 & 0 & 133 & 0 & 0 & -134 & 0 \\
 0 & 0 & 0 & 0 & 24 & 0 & 0 & 144 & 0 & 0 & -1064 \\
 0 & 0 & -3 & 0 & 0 & 136 & 0 & 0 & -871 & 0 & 0 \\
 0 & 0 & 0 & 12 & 0 & 0 & -134 & 0 & 0 & -1292 & 0 \\
 0 & 0 & 0 & 0 & 27 & 0 & 0 & -1064 & 0 & 0 & 6072
\end{array}
\right)
\end{equation}

Now we would like to match the predictions of equation (\ref{PLg3con}) with the gauge invariant operators in quiver theory
\begin{eqnarray}\label{n3}
4t_1^3 & \rightarrow & \epsilon \epsilon A_i A_j A_k  \hbox{ :         4 operators} \nonumber \\
2t_1^4t_2 & \rightarrow & ... \hbox{ :        2 non factorizable baryons } \nonumber \\
2t_1 t_2^4 & \rightarrow & ... \hbox{ :       2 non factorizable baryons }\nonumber\\
4t_2^3 & \rightarrow & \epsilon \epsilon B_i B_j B_k  \hbox{ :       4 operators } \nonumber\\
4 t_1 t_2 & \rightarrow &  \hbox{Tr}(A_i B_j) \hbox{ :       4 operators } \nonumber\\
9t_1^2t_2^2 & \rightarrow & \hbox{Tr}(A_i B_j A_k B_l) \hbox{ :        9  operators} \nonumber \\
\end{eqnarray}
As at the level $N=2$ the operators of the type $\hbox{Tr}(ABABAB)$ do not appear independently but are generated by the basic operators in (\ref{n3}). This can be understood in the same way as the case $N=2$ using the tensor relation (\ref{tensrel}). 

The most interesting basic generators are the ones related to the monomials $2t_1^4 t_2$, $2t_1t_2^4$: these are the first cases of non factorizable baryons. Let us analyze the monomial $2t_1^4 t_2$ (the discussion for the other monomial is the same if one exchanges the $A$ fields with the $B$ fields and $t_1$ with $t_2$). At this level, if all the baryonic operators were factorizable, the only gauge invariant operators we could construct with the right quantum numbers would be of the type $(\epsilon \epsilon A_i A_j A_k)(\hbox{Tr} A_l B_m)$ and there are just $16$ of them. If instead we follow the general proposal \cite{Beasley:2002xv,Butti:2006au } to relate monomials in the homogeneous coordinates to baryonic gauge invariant operators, at the level $t_1^4 t_2$ we would write the operators

\begin{equation}\label{nonfact}
\epsilon \epsilon (A_lB_mA_i) (A_j) (A_k),
\end{equation}
where the epsilon terms are contracted once with each index in the brackets.
If we count how many operators exist of this form we discover that they are $18$. This suggests that among the operators in (\ref{nonfact}) there are $2$ that are not factorizable and must be included as generators of the ring. An explicit
computation shows that there are indeed two non factorizable determinants.
These two ``special'' baryons in the spectra of the field theory are the ones that make the generating function $g_3(t_1,t_2;{\cal C})$ non trivial, and they are precisely the ones required by the plethystic logarithm (\ref{PLg3con}) as basic generators at the level $t_1^4 t_2$. The existence of non factorizable baryons is related
to the flavor indices of the $A_i$ \cite{Berenstein:2002ke}; 
only those baryons where the $A_i$ are
suitably symmetrized can be factorized using the relation (\ref{tensrel}).

Next, one can go up with the values of $N$ and try to extract the general pattern of basic generators for the chiral ring of the gauge theory
dual to the conifold singularity.

We use a notation where $( . , . )$ denotes the weights of the gauge invariant operators under $t_1$, $t_2$. 
The basic generators for the first few values of $N$ are: 
\begin{eqnarray}\label{NNN}
N=1 &\rightarrow& 2 (1,0) + 2 (0,1) \hbox{ :   4 operators} \nonumber\\ 
N=2 &\rightarrow& 3 (2,0) + 4 (1,1) + 3 (0,2)   \hbox{ :  10 operators} \nonumber\\  
N=3 &\rightarrow& 4 (3,0) + 2 (4,1) + 4 (1,1) + 9 (2,2) +  \nonumber\\ & &  2 (1,4) + 4 (0,3)   \hbox{ :  25 operators} \nonumber\\     
N=4 &\rightarrow& 5 (4,0) + 4 (5,1) + 4 (1,1) + 9 (2,2) + 16 (3,3) +  \nonumber\\ & & 4 (1,5) + 5(0,4)   \hbox{ :  47 operators} \nonumber\\ 
N=5 &\rightarrow& 6 (5,0) + 6 (6,1) + 6 (7,2) + 4 (1,1) + 9 (2,2) + 16 (3,3) + 25 (4,4) + \nonumber\\ & & 6 (2,7)+ 6 (1,6)+6 (0,5)   \hbox{ :  90 operators} 
\end{eqnarray}
From (\ref{NNN}) it seems that the pattern of the basic generators for generic $N$ is the following (we assume reflection symmetry under the exchange of $t_1$ and $t_2$):
\begin{itemize}
\item{The mesonic generators of the chiral ring are all the mesons (single traces) starting from weight (1,1) up to weight $(N-1,N-1)$. There are $(n+1)^2$ generators with weight $(n,n)$, transforming in the spin $(\frac{n}{2},\frac{n}{2})$ representation, leading to a total of $\frac{(N-1)(2N^2+5N+6)}{6}$ generators;}
\item{The baryonic generators of the chiral ring have a jump of 2 in the level:
\begin{itemize}
\item{$N+1$ generators of weight $(N,0)$;}
\item{$2(N-2)$ generators of weight $(N,1)$ starting at level $N=3$, which are exactly the non factorizable baryons we have already discussed;}
\item{generators of weight $(N,2)$ starting at level $N=5$, which are new non factorizable baryons;}
\item{etc.}
\end{itemize}}\end{itemize}
The non factorizable baryons appear for the first time at level $N=3$, and going up with the number of branes the number and type of non trivial baryons increase.

\subsection{A Toy Model -- Half the Conifold}

\begin{figure}[h!!!!!!!!!!!!!!!!!]
\begin{center}
\includegraphics[scale=0.5]{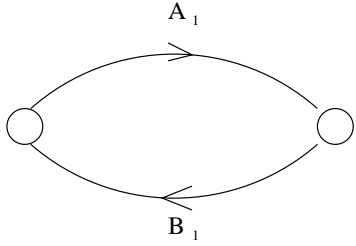} 
\caption{Quiver for half the conifold.}
\label{half}
\end{center}
\end{figure}

In this subsection we are going to look at a toy model which consists of half the matter content of the conifold. This model is particularly simple as it has no F term relations and all baryons are factorized. 
The gauge theory data can be encoded in the quiver in Figure \ref{half} with $W=0$.

The chiral multiplets are assigned charges according to Table \ref{globalhalfconifold} which is a reduction of Table \ref{globalconifold}.

\begin{table}[htdp]
\caption{Global charges for the basic fields of the quiver gauge theory for ``Half the Conifold."}
\begin{center}
\begin{tabular}{|c||c|c||c|}
\hline
& $U(1)_R$ & $U(1)_B$ & monomial\\ \hline
$A_1$ & $\frac{1}{2}$ & 1& $t_1 = t b$\\ \hline
$B_1$ & $\frac{1}{2}$ & -1& $t_2 = \frac{t}{b}$ \\ \hline
\end{tabular}
\end{center}
\label{globalhalfconifold}
\end{table}%

With this toy model we are counting the subset of BPS operators of the conifold
with no occurrences of $A_2$ and $B_2$.

As for the conifold case, the generating function for one D brane and any baryon number is freely generated by the two chiral multiplets $A_1$ and $B_1$ and takes the form

\begin{equation}
g_1(t_1,t_2; \frac{1}{2}{\cal C}) = \frac{1}{(1-t_1)(1-t_2)} = \frac{1}{(1-t b) (1- \frac {t}{b})}.
\label{g1halfconi}
\end{equation}

From which it is easy to extract the generating function for fixed baryonic charge $B$ and one D brane, $N=1$,

\begin{eqnarray}
g_{1,B\ge0}(t_1, t_2; \frac{1}{2}{\cal C}) &=& \frac{t_1^B} {1-t_1t_2} = \sum_{n=0}^\infty t_1^{n+B} t_2^n, \nonumber \\ \nonumber
g_{1,B\le0}(t_1, t_2; \frac{1}{2}{\cal C}) &=& \frac{t_2^{-B}} {1-t_1t_2} = \sum_{n=0}^\infty t_1^{n} t_2^{n-B}.
\label{reshalfcon}
\end{eqnarray}
which indeed reflects the symmetry of taking $B\leftrightarrow -B$, $t_1\leftrightarrow t_2$ simultaneously.
Setting $\nu$ to be the chemical potential for the number of D branes, $N$, and taking the $\nu$-inserted plethystic exponential of these expressions we can easily extract the the generating function for fixed number of baryons

\begin{eqnarray}
g_{B}(\nu; t_1, t_2; \frac{1}{2}{\cal C}) &=& \exp\biggl(\sum_{k=1}^\infty \frac{\nu^k}{k}g_{1,B}(t_1^k, t_2^k; \frac{1}{2}{\cal C}) \biggr), \nonumber \\ \nonumber
g_{B\ge0}(\nu; t_1, t_2; \frac{1}{2}{\cal C}) &=& \exp\biggl(\sum_{k=1}^\infty \frac{\nu^k t_1^{B k}}{k (1-t_1^k t_2^k)} \biggr), \nonumber \\ \nonumber
g_{B\le0}(\nu; t_1, t_2; \frac{1}{2}{\cal C}) &=& \exp\biggl(\sum_{k=1}^\infty \frac{\nu^k t_2^{ - B k}}{k (1-t_1^k t_2^k)} \biggr).
\end{eqnarray}
We can compare this generating function to the generating function of the complex line, taken from \cite{Benvenuti:2006qr},

\begin{equation}
g(\mu ; t ; \mathbb{C}) = \exp\biggl(\sum_{k=1}^\infty \frac{\mu^k}{k(1-t^k)}\biggr) = \sum_{N=0}^\infty \mu^N \prod_{i=1}^N \frac{1}{(1-t^i)} = \prod_{k=0}^\infty \frac{1}{(1-\mu t^k)},
\end{equation}
which upon substitution $\mu=\nu t_1^B$ and $t=t_1 t_2$ for positive baryonic charge and similar expressions for negative baryonic charge, leads to the generating function
for fixed number of D branes $N$ and fixed baryonic charge $N B$,

\begin{eqnarray}
g_{N,N B\ge0}(t_1, t_2; \frac{1}{2}{\cal C}) &=& t_1^{NB}\prod_{i=1}^{N}\frac{1} {(1-t_1^it_2^i)}, \nonumber \\ \nonumber
g_{N,N B\le0}(t_1, t_2; \frac{1}{2}{\cal C}) &=& t_2^{-NB}\prod_{i=1}^{N}\frac{1} {(1-t_1^it_2^i)}.
\end{eqnarray}

Note the relation  $g_{N,N B\ge0}=t_1^{NB}g_{N,0}$ (and a similar expression for $B<0$) which implies that the partition function at baryonic number $N B$ is
proportional to the mesonic partition function. This can be understood
by observing that, since there are no flavor indices, using equation (\ref{tensrel}), we can factorize all baryons into a product of the basic 
baryons $\det A_1$ and $\det B_1$ times mesons. 

Summing over all baryonic charges we can get the generating function for fixed number of D3-branes, $N$

\begin{equation}
g_N(t_1,t_2; \frac{1}{2}{\cal C}) = \frac{1}{(1-t_1^N)(1-t_2^N)} \prod_{i=1}^{N-1}\frac{1} {(1-t_1^it_2^i)}.
\label{gNhalfconi}
\end{equation}

Which turns out to be freely generated. The corresponding chiral ring of BPS operators is generated by the $N+1$ operators, $\det A_1$, $\det B_1$, and Tr $(A_1B_1)^i, i=1\ldots N-1$.

\subsubsection{Comparing with the Molien Invariant}

The generating functions for cases in which the superpotential is zero can be computed using a different method -- that of the Molien Invariant. This method was used independently by Pouliot \cite{Pouliot:1998yv} and by R\"omelsberger \cite{Romelsberger:2005eg} to compute generating functions for BPS operators in the chiral ring for a selected set of gauge theories with zero superpotential, $W=0$.
It is using a multi-contour integral formula and goes as follows. Given a supersymmetric ${\cal N} = 1$ gauge theory with a gauge group $G$ of rank $r$ and Weyl group of order $|W|$, and a set of chiral multiplets transforming in representations, $R_k$, we set a chemical potential $t_k$ for each representation and compute the generating function for BPS operators in the chiral ring

\begin{equation}
g(\{t_k\}; G) = \frac{1}{|W|} \prod_{j=1}^r \oint \frac{dw_j}{2 \pi i w_j} \frac{\prod_\alpha(1-w^{h(\alpha)})}{\prod_k \prod_{\lambda_k}(1- t_k w^{h(\lambda_k)})} ,
\end{equation}
where $h(\alpha)$ are weights of the adjoint representation of $G$, while $h(\lambda_k)$ are weights of the representation $R_k$.

This function is also used in the Commutative Algebra literature (see for example \cite{Djokovic:1981bh}) and is called there the Molien--Weyl invariant.

For the case of a gauge group $SU(N)$ and $n$ chiral multiplets in the fundamental representation we introduce $n$ chemical potentials, $t_k, k=1\ldots n$ and this multi-contour integral can be extended to an $N$-dimensional contour integral of the form

\begin{equation}
g(\{t_k\}; SU(N) ) = \frac{1}{N!} \prod_{j=1}^N \oint \frac{dw_j}{2 \pi i w_j} \frac{\prod_{i<j} (w_i-w_j)^2}{1-\prod_{j=1}^N w_j} \frac {1}{\prod_{k=1}^n \prod_{j=1}^N (1- t_k w_j)} .
\end{equation}

This expression counts all gauge invariant operators which contain fields in the fundamental representation but not fields in the anti-fundamental representation.
If we would like to count BPS operators which are constructed out of $n_1$ chiral multiplets in the fundamental representation and $n_2$ chiral multiplets in the anti-fundamental representation we introduce chemical potentials $t_{k_1}, k_1=1\ldots n_1$ and $q_{k_2}, k_2=1\ldots n_2$ and write
the integral

\begin{eqnarray}
& & g(\{t_{k_1}\}, \{q_{k_2}\}; SU(N) ) = \\ \nonumber
& & \frac{1}{N!} \prod_{j=1}^N \oint \frac{dw_j}{2 \pi i w_j}
\frac{\prod_{i<j} (w_i-w_j)^2}{1-\prod_{j=1}^N w_j}
\frac {1}{\prod_{k_1=1}^{n_1} \prod_{j=1}^N (1- t_{k_1} w_j)}
\frac {1}{\prod_{k_2=1}^{n_2} \prod_{j=1}^N (1- t_{k_2} w_j^{-1})}.
\end{eqnarray}

We are now ready to present the formula for the case of discussion in this section, half the conifold. We have two gauge groups, $SU(N)\times SU(N)$ for which we introduce 2 sets of $N$ variables each, $w_i, i=1\ldots N$, and $v_i, i=1\ldots N$. There is one chiral multiplet $A_1$ with a chemical potential $t_1$ and one chiral multiplet $B_1$ with 

\begin{eqnarray}
& & g_N(t_1, t_2; \frac{1}{2}{\cal C}) = \\ \nonumber
& & \frac{1}{(N!)^2}
\prod_{r=1}^N \oint \frac{dw_r}{2 \pi i w_r}
\prod_{s=1}^N \oint \frac{dv_s}{2 \pi i v_s}
\frac{\prod_{i<j} (w_i-w_j)^2 (v_i-v_j)^2}{(1-\prod_{j=1}^N w_j) (1-\prod_{j=1}^N v_j)}
\frac {1}{\prod_{r,s=1}^N (1- \frac{t_1 w_r}{v_s})(1- \frac{t_2 v_s}{w_r})}.
\end{eqnarray}

The case $N=2$ was computed explicitly and the result coincides with Equation (\ref{gNhalfconi}), while other cases were not checked due the heavy computations they involve but definitely can be checked for higher values of $N$.

\subsection{Another Example -- $\frac{3}{4}$ Conifold}

Here we are considering another example of counting BPS operators which are a subset of the states in the conifold theory. We are going to restrict to all the operators that include the fields $A_1, A_2, B_1$ in Table \ref{globalconifold} but not $B_2$. Since these are 3 out of 4 of the fundamental fields of the conifold theory we use the suggestive name ``$\frac{3}{4}$ Conifold". It is amusing to see that some of the expressions can be derived exactly and for any number of branes $N$ and for any baryonic number $B$, even though the full answer can be represented implicitly but no explicit expression is known.
The first point to note is that the field $B_1$ has no flavor index and therefore all baryons with negative baryonic charge factorize. This allows us to sum exactly all the generating functions with negative baryonic charge and fixed $N$. The F term relations are indeed relevant for this problem but they turn out to affect the mesons only ($B=0$). For simplicity of computations we are going to set $x=1$ while the chemical potential $y$ for the second $SU(2)$ is not needed.

The $N=1$ generating function is again freely generated and is given by equation (\ref{g1coni}) with the factor corresponding to $B_2$ removed. It takes the form 

\begin{equation}
g_1(t_1,t_2; \frac{3}{4}{\cal C}) = \frac{1}{(1-t_1)^2(1-t_2)} = \frac{1}{(1-t b)^2 (1- \frac {t}{b})}.
\label{g134coni}
\end{equation}

Using techniques which by now are standard (see equations (\ref{res1}, \ref{res2}, \ref{res3})) we have

\begin{eqnarray}
g_{1,B\ge0}(t_1, t_2; \frac{3}{4}{\cal C}) &=& \frac{t_1^B (1+B - B t_1 t_2)} {(1-t_1t_2)^2} = \sum_{n=0}^\infty (n+B+1) t_1^{n+B} t_2^n, \nonumber \\
g_{1,B\le0}(t_1, t_2; \frac{3}{4}{\cal C}) &=& \frac{t_2^{-B}} {(1-t_1t_2)^2} = \sum_{n=0}^\infty (n+1) t_1^{n} t_2^{n-B}.
\end{eqnarray}
and applying the plethystic exponential we find the generating function for fixed $B$.

\begin{eqnarray}
\nonumber
g_{B\ge0}(\nu; t_1, t_2; \frac{3}{4}{\cal C}) &=& \exp\biggl(\sum_{k=1}^\infty \frac{\nu^k t_1^{ B k}}{k (1-t_1^k t_2^k)^2} \biggr) \biggl[ \exp\biggl(\sum_{k=1}^\infty \frac{\nu^k t_1^{B k}}{k (1-t_1^k t_2^k)} \biggr) \biggr]^B \\ \nonumber
g_{B\le0}(\nu; t_1, t_2; \frac{3}{4}{\cal C}) &=& \exp\biggl(\sum_{k=1}^\infty \frac{\nu^k t_2^{ - B k}}{k (1-t_1^k t_2^k)^2} \biggr) = \prod_{n=0}^\infty \frac{1}{(1 - \nu t_1^n t_2^{n-B})^{n+1}}
\end{eqnarray}

We notice that, as in the half-conifold case, 

\begin{equation}
g_{B\le0}(\nu; t_1, t_2; \frac{3}{4}{\cal C}) = g_{B=0}(\nu t_2^{-B}; t_1, t_2; \frac{3}{4}{\cal C}) ,
\end{equation}

This implies an equality satisfied by generating functions with fixed number of branes and fixed negative baryonic charge by taking the power expansion in $\nu$ on both sides,

\begin{equation}
g_{N,N B\le0}(t_1, t_2; \frac{3}{4}{\cal C}) = t_2^{- N B} g_{N,0}(t_1, t_2; \frac{3}{4}{\cal C}) ,
\end{equation}

Next we sum over all negative baryonic charges and get

\begin{equation}
\sum_{B=0}^\infty t_2^{N B} g_{N,0} (t_1, t_2; \frac{3}{4}{\cal C}) = \frac{g_{N,0}(t_1, t_2; \frac{3}{4}{\cal C})}{1-t_2^N}
\end{equation}
so that just one free generator ($\det B_1$) is added to the mesonic ones.
In agreement with the fact that baryons are factorized and relations only
apply to mesons.

Notice that the partition function for mesons, $g_{B=0}$ with $t_1 t_2=t$
is the mesonic partition function for $\mathbb{C}^2$, namely

\begin{equation}
\nonumber
g_{B=0}(\nu; t_1, t_2; \frac{3}{4}{\cal C}) = \exp\biggl(\sum_{k=1}^\infty \frac{\nu^k}{k (1-t_1^k t_2^k)^2} \biggr) = g(\nu; t_1 t_2 ; \mathbb{C}^2) 
\end{equation}

This follows from the fact that the two matrices $A_1 B_1$ and $A_2 B_1$ are adjoint valued and
commute due to F-term relations. As a result their moduli space is a copy of $\mathbb{C}^2$.

The other half of the spectrum with $\det A_i$ is more difficult to analyze:
there are non factorizable baryons and non trivial relations. To get some expressions for the positive case we write 

\begin{eqnarray}
\nonumber
g_{N B\ge0}(\nu; t_1, t_2; \frac{3}{4}{\cal C}) &=& \biggl[ \sum_{N=0}^\infty \nu^N t_1^{N B} g_{N,0}(t_1, t_2; \frac{3}{4}{\cal C})\biggr] \biggl[ \sum_{N=0}^\infty \nu^N t_1^{N B} g_N(t_1 t_2 ; \mathbb{C}) \biggr]^B \end{eqnarray}

The baryon number dependence is indeed isolated but does not seem to simplify to be summed over.

\subsubsection{$N=2$ for $\frac{3}{4}$ the Conifold}

Getting a general expression for any $N$ seems to be too hard. We can nevertheless use the techniques described in Section \ref{Nis2} in order to get an explicit expression for $N=2$. Using equations (\ref{g2coni}) and (\ref{g134coni}), we find

\begin{eqnarray}
\nonumber
g_2(t_1, t_2; \frac{3}{4} {\cal C}) &=& \frac{1}{2}g_1(t_1^2,t_2^2; \frac{3}{4} {\cal C})
+ \frac{1}{2}\frac{1}{(2\pi i)} \oint \frac {ds} {s} g_1(t_1 s,\frac{t_2} {s}; \frac{3}{4} {\cal C})  g_1(\frac{t_1}{s},t_2 s; \frac{3}{4} {\cal C}) \\
&=& \frac{1-t_1^4 t_2^2}{(1-t_1^2)^3(1-t_1 t_2)^2(1-t_2^2)},
\label{g234coni}
\end{eqnarray}
implying that the moduli space of vacua for this theory is a complete intersection and is generated by the 6 operators, the spin 1 baryons, $\det A_1, \epsilon \epsilon A_1 A_2, \det A_2$, the spin $\frac{1}{2}$ mesons $Tr (A_1 B_1), Tr (A_2 B_1)$, and the spin 0 baryon $\det B_1$. The moduli space is five dimensional, and is given by a degree (4,2) (4 $A$'s, 2 $B$'s) relation in these 6 generators.

\section{Baryonic Generating Functions for the Orbifold $  \frac{\mathbb{C}^2}{Z_2} \times \mathbb{C} $}
\label{secC2Z2}
\setall

\begin{figure}[h!!!]
\begin{center}
\includegraphics[scale=0.5]{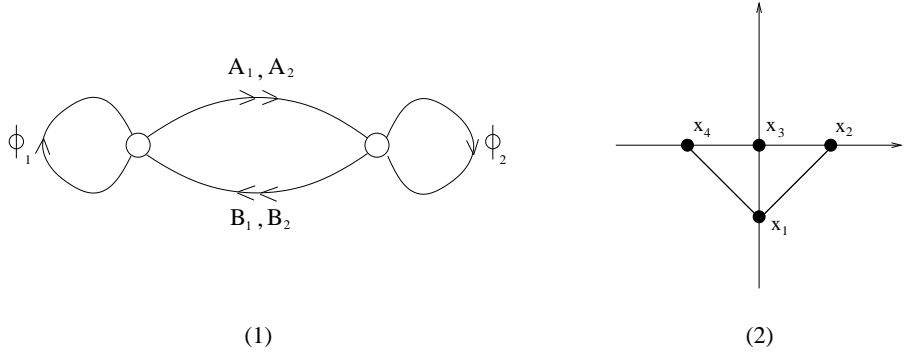} 
\caption{The quiver (1) and the toric diagram (2) for the $ \frac{\mathbb{C}^2}{Z_2} \times \mathbb{C} $ singularity.}\label{C2Z2}
\end{center}
\end{figure}

The quiver gauge theory living on a D brane probing the orbifold $  \frac{\mathbb{C}^2}{Z_2} \times \mathbb{C} $ is depicted in figure \ref{C2Z2}. 

It has ${\cal N}=2$ supersymmetry with two vector multiplets in the adjoint representation of $SU(N)$ and 2 bi-fundamental hypermultiplets. In ${\cal N}=1$ notation we have 6 chiral multiplets denoted as $\phi_1, \phi_2, A_1, A_2, B_1, B_2$, with a superpotential

\begin{equation}
W = \phi_1 (A_1 B_2 - A_2 B_1) + \phi_2 (B_1 A_2 - B_2 A_1)
\end{equation}

The global symmetry for this theory is $SU(2)_R\times U(1)_3\times U(1)_R\times U(1)_B$ and the corresponding charges under this global symmetry are given in Table \ref{globalC2Z2}.

\begin{table}[htdp]
\caption{Global charges for the basic fields of the quiver gauge theory on the D brane probing the orbifold $  \frac{\mathbb{C}^2}{Z_2} \times \mathbb{C} $.}
\begin{center}
\begin{tabular}{|c||c c|c|c|c||c|}
\hline
 & \multicolumn{2} {c|} {$SU(2)_R$} 		& {$U(1)_3$}	& $U(1)_R$	& $U(1)_B$ & monomial\\ \hline
\cline{2-5}
		& $j_1$		& $m_1$		&			&			&&\\ \hline \hline
$A_1$ 	& $\frac{1}{2}$	& $+\frac{1}{2}$& 0	& $\frac{2}{3}$	& 1	& $t_1 x$\\ \hline
$A_2$ 	& $\frac{1}{2}$	& $-\frac{1}{2}$	& 0	& $\frac{2}{3}$	& 1	& $\frac{t_1}{x}$ \\ \hline
$B_1$ 	& $\frac{1}{2}$	& $+\frac{1}{2}$& 0	& $\frac{2}{3}$	& -1	& $t_2 x$ \\ \hline
$B_2$ 	& $\frac{1}{2}$	& $-\frac{1}{2}$	& 0	& $\frac{2}{3}$	& -1	& $\frac{t_2}{x}$ \\ \hline
$\phi_1$ 	& 0			& 0			& 2	& $\frac{2}{3}$	& 0	& $t_3$ \\ \hline
$\phi_2$ 	& 0			& 0			& 2	& $\frac{2}{3}$	& 0	& $t_3$ \\ \hline
\end{tabular}
\end{center}
\label{globalC2Z2}
\end{table}%

Note that both $U(1)_R$ and $U(1)_3$ are R symmetries but $U(1)_3$ is anomalous while $U(1)_R$ is anomaly free. The mesonic generating function on the Higgs branch was computed in \cite{Benvenuti:2006qr} and takes the form

\begin{equation}
g_{B=0}(\nu ; t_1, t_2, t_3 ;   \frac{\mathbb{C}^2}{Z_2} \times \mathbb{C}  ) =  \exp\biggl(\sum_{k=1}^\infty\frac{\nu^k}{k} \frac{1+t_1^{k} t_2^{k}}{(1-t_1^{2k})(1-t_2^{2k})(1-t_3^k)} \biggr) ,
\label{Hnu}
\end{equation}

The generating function on the mixed Coulomb and Higgs branch was computed in \cite{Hanany:2006uc}.
Both these works do not count baryonic operators. In particular there is no counting for gauge invariant operators in which the number of $A$'s is different than the number of $B$'s.
For simplicity we will not count operators with adjoint fields and therefore mostly going to ignore the $\mathbb{C}$ factor in the following expressions.
We will now proceed to compute the baryonic generating function.
As before we start by analysis of the $N=1$ case.

\subsection{$N=1$ for $\frac{\mathbb{C}^2}{Z_2}$} 

The case $N=1$ for the orbifold is slightly different from that for the conifold.
The reason is that, even for $N=1$, the superpotential does not vanish. In
particular there is a non trivial F term relation $A_1 B_2= A_2 B_1$.
There are 3 global charges for this case and 4 fields with one relation.
This leads to expectation that the $N=1$ generating function is a complete intersection.
We can also write an expression which will take care of the $\mathbb{C}$ factor,

\begin{equation}
g_1(t_1,t_2,t_3,x;  \frac{\mathbb{C}^2}{Z_2} \times \mathbb{C}  )= \frac{1}{1-t_3} g_1(t_1,t_2,x;  \frac{\mathbb{C}^2}{Z_2}   )
\label{tota}
\end{equation}
and write the generating function of the $\frac{\mathbb{C}^2}{Z_2}$ part as
generated by four fields subject to one relation,

\begin{eqnarray}
\nonumber
g_1(t_1,t_2,x;  \frac{\mathbb{C}^2}{Z_2} ) &=& \frac{1-t_1 t_2}{(1- t_1 x)(1 - \frac{t_1}{x})(1-t_2 x)(1- \frac{t_2}{x})} \\
&=& \frac{1-t^2}{(1- t b x)(1 - \frac{t b}{x})(1-\frac{t x}{b})(1- \frac{t}{b x})} .
\label{g1C2Z2}
\end{eqnarray}

This generating function describes a moduli space which is a 3 dimensional complete intersection, generated by the chiral multiplets $A_1, A_2, B_1, B_2$, which satisfy one relation $A_1 B_2 = A_2 B_1$.

Following the same procedure as in section \ref{conifold} for the conifold theory we find

\begin{eqnarray}
g_{1,B\ge0}(t,x;  \frac{\mathbb{C}^2}{Z_2} )
&=& \frac{t^B x^{B} } { (1 - \frac{1}{x^2}) (1-t^2 x^2)}
+ \frac{t^B x^{-B}} { (1 - x^2)(1-\frac{t^2}{x^2}) } \nonumber \\
g_{1,B\le0}(t,x;  \frac{\mathbb{C}^2}{Z_2} ) &=& \frac{t^{-B} x^{-B} } { (1-\frac{1}{x^2}) (1-t^2 x^2) }
+ \frac{t^{-B} x^{B}} { (1-x^2)(1-\frac{t^2}{x^2}) } .
\label{g1BC2Z2}
\end{eqnarray}

\begin{figure}[h!!!]
\begin{center}
\includegraphics[scale=0.6]{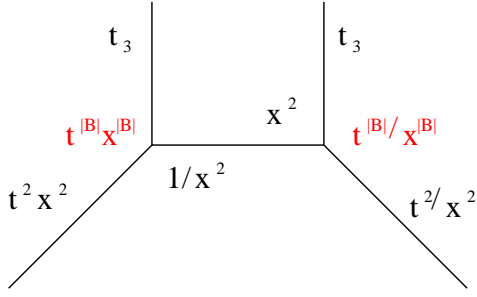} 
\caption{The generating functions $g_{1,B}$ can be computed using the index
theorem and  written as sum over fixed points corresponding to  the $(p,q)$ web vertices of $\frac{\mathbb{C}^2}{Z_2} \times \mathbb{C}$. 
As in the conifold case, localization needs to be performed in a correctly
normalized basis of charges for $T^3$, which are related to $(t_3,x,t)$ by
equation (\ref{rel}).
The vertical legs contribute the factorized factor $1/(1-t_3)$ in equation (\ref{tota}).}
\label{diagramZ2toric}
\end{center}
\end{figure}

These generating functions can be described diagrammatically by the toric diagram of the $\frac{\mathbb{C}^2}{Z_2} \times \mathbb{C}$ space, as depicted in Figure \ref{diagramZ2toric}.

By comparing Equation (\ref{g1C2Z2}) with Equation (\ref{chebymesonid}) we find that $g_1$ admits a Chebyshev polynomial expansion, with the assignments $x=e^{i \theta}, b=e^{i \beta}$,

\begin{eqnarray}
g_{1}(t,b, x;  \frac{\mathbb{C}^2}{Z_2}   )
&=& \sum_{n=0}^\infty t^{n} U_{n} (\cos \theta) U_{n} (\cos \beta) \\ \nonumber
&=& \frac{1-t^2} {(1-t b x) (1-\frac{t b}{x}) (1-\frac{t x} {b}) (1-\frac{t}{b x}) } .
\label{chebyC2Z2}
\end{eqnarray}

This is a surprising point and suggests that the baryonic charge is a $U(1)$ Cartan subgroup of a hidden global $SU(2)_H$ symmetry. This symmetry can not be present in the classical theory since it acts differently on the $A$'s and $B$'s which in turn do not carry the same gauge quantum numbers. We will come back to this point for higher number of D branes. 

\subsection{Comparison with the Geometric Generating Function}
We make a digression to explore the relation of the $N=1$ generating function 
of BPS gauge invariant operators with
the natural geometric counterpart, the generating function for homogeneous
polynomials. We saw that, in the case of the conifold, the two generating
functions coincide.
In this respect, the conifold is unique since there is a 
one-to-one correspondence between homogeneous coordinates and elementary fields.
For a generic toric cone, although, modulo Seiberg's dualities, the quiver is completely determined by
the toric data, the relation between homogeneous coordinates and fields
is more involved. 


The toric diagram for $\frac{\mathbb{C}^2}{Z_2} \times \mathbb{C}$ is shown
in Figure \ref{C2Z2}. There are four homogeneous coordinates associated
with the four integer points on the perimeter of the toric diagram.
We choose the following relation between the homogeneous charges and
the baryonic and flavor charges 
\begin{equation} x_1=t_3 \, \qquad x_2= b t x \, \qquad x_3=\frac{1}{b^2}\, \qquad x_4= \frac{b t}{x} \end{equation}
where $b$ denotes as before the baryonic charge. The assignment of baryonic
charge is compatible with the general formula (\ref{bartoric}).

It is a simple exercise in tiling construction to assign homogeneous charges
to the elementary fields \cite{kru2,Butti:2005vn}. 
The fields $A_i$ with baryonic charge one are associated with 
$x_2,x_4$.
The fields $B_i$ with baryonic charge minus one 
are associated instead with $x_2 x_3$ and $x_3 x_4$. 
In terms of global charges, $b$ counts the baryonic charge, $x$ counts the $SU(2)$ numbers and $t$ the dimension. 
Notice that the coordinate $x_3$, with
baryonic charge $-2$ is not associated with an elementary field. 

The generating function for homogeneous polynomials is
\begin{equation}
\frac{1}{(1-t_3)(1-b t x)(1- \frac{b t}{x})(1-\frac{1}{b^2})}=\frac{1}{1-t_3} \sum_{-\infty}^{\infty} b^B Z_B (x,t)
\end{equation}
The factor $\frac{1}{(1-t_3)}$ is factorized and will be ignored in the following.
The functions $Z_B$ enumerate homogeneous monomials with baryonic charge $B$
and we could expect that $Z_B$ coincides with the previously computed
$g_{1,B}(t,x;  \frac{\mathbb{C}^2}{Z_2})$. 
This is actually true for $B\ge -1$, as can be explicitly checked
by expanding the generating function 
in Laurent series, but fails for $B\le -2$. The reason is that geometry
at baryonic number $B=-2$ counts all homogeneous monomials with $B=-2$
starting with $x_3$. Since $x_3$ is not associated with a field, these
geometrical objects have no direct relation with field theory.

We see that the expansion of the geometric generating function contains 
all the basic ingredients $g_{1,B}$ which serve for the construction of 
the quantum field theory partition functions but packed together in a
different way. The field theory generating function can be reconstructed
from the geometric one by using the following observations: 
the positive baryonic
charge partition functions $Z_B$ give the correct result, computing 
BPS operators with more $A$'s than $B$'s,  and moreover the
orbifold has an obvious $Z_2$ symmetry between $A_i$ and $B_i$.
So for $N=1$ the natural expectation for the field theory generating function
is
\begin{equation} \sum_{B=1}^\infty b^B Z_B +  \sum_{B=1}^\infty b^{-B} Z_B + Z_0=\frac{1-t^2}
{(1-b x t)(1- \frac{b t}{x})(1- \frac{t x}{b})(1- \frac{t}{b x})}
\nonumber
\end{equation}
which indeed coincides with equation~(\ref{g1C2Z2}).

\subsection{$N=2$ for the Orbifold $ \frac{\mathbb{C}^2}{Z_2}$}
Using methods by now familiar we compute

\begin{eqnarray}\label{g2C2Z2}
& & g_2(t_1,t_2,x;  \frac{\mathbb{C}^2}{Z_2}) \\
&=&\frac{\left(1-t_1^2 t_2^2\right)
   [(1-t_1^2 t_2^2 x^2) (1- \frac{t_1^2 t_2^2} {x^2})+t_1t_2(1-t_1^2)(1-t_2^2)]}
   {\left(1-t_1^2\right)
   \left(1-t_2^2\right) (1-t_1^2 x^2)(1- \frac{t_1^2} {x^2}) (1- t_1 t_2 x^2)(1- \frac{ t_1 t_2} {x^2})
  (1-t_2^2 x^2)(1- \frac{t_2^2} {x^2})
}\nonumber
\end{eqnarray}

On the $(t_1,t_2)$ lattice the first few terms look like

\begin{equation}
\left(
\begin{array}{lllllllll}
 1 & 0 & 3 & 0 & 6 & 0 & 10 & 0 & 15 \\
 0 & 3 & 0 & 8 & 0 & 15 & 0 & 24 & 0 \\
 3 & 0 & 11 & 0 & 22 & 0 & 36 & 0 & 53 \\
 0 & 8 & 0 & 22 & 0 & 40 & 0 & 62 & 0 \\
 6 & 0 & 22 & 0 & 45 & 0 & 73 & 0 & 106 \\
 0 & 15 & 0 & 40 & 0 & 73 & 0 & 112 & 0 \\
 10 & 0 & 36 & 0 & 73 & 0 & 119 & 0 & 172 \\
 0 & 24 & 0 & 62 & 0 & 112 & 0 & 172 & 0 \\
 15 & 0 & 53 & 0 & 106 & 0 & 172 & 0 & 249
\end{array}
\right)
\end{equation}

Taking the Plethystic Logarithm of Equation (\ref{g2C2Z2}) we find

\begin{eqnarray}
\nonumber
f_2(t_1,t_2,x;  \frac{\mathbb{C}^2}{Z_2} ) &=& \chi_1(x) (t_1^2+t_1 t_2+t_2^2) - t_1 t_2^3 - t_2 t_1^3 - (\chi_0(x) + \chi_1(x) ) t_1^2t_2^2 + \ldots \\
&=& t^2 \chi_1(x) \chi_1(b) - t^4 ( \chi_1(b) + \chi_1(x) ) + \ldots
\end{eqnarray}
implying that there are nine generators, quadratic in the basic fields of table \ref{globalC2Z2} transforming in the spin (1,1) representation of $SU(2)_R\times SU(2)_H$. They satisfy 6 relations, quartic in the basic fields and hence quadratic in the generators, that transform as spin (1,0) + (0,1) of the global symmetry. Explicitly, the generators are the following objects, conveniently arranged into a $3\times3$ matrix:

\[ \left( \begin{array}{ccc}
\det A_1 & \epsilon \epsilon A_1 A_2 & \det A_2 \\
\hbox{ Tr($A_1B_1$) } & \hbox{ Tr($A_1 B_2$)} & \hbox{ Tr($A_2 B_2$)} \\
\det B_1 & \epsilon \epsilon B_1 B_2 & \det B_2 \end{array} \right)\]

Note that $A_1 B_2 = A_2 B_1$ by the F term relations and therefore only one of them needs to be counted. There are two relations among the operators $\epsilon \epsilon A A \hbox{ Tr($A B$) }$ and  
$\epsilon \epsilon B B \hbox{ Tr($A B$) }$ and four \
among $\epsilon \epsilon A A \epsilon \epsilon B B$.

One can analogously compute the $N=3$ generating function. The expression is too long to present here but we can show the first few terms in the $(t_1,t_2)$ lattice,

\begin{equation}
\left(
\begin{array}{lllllll}
 1 & 0 & 0 & 4 & 0 & 0 & 10 \\
 0 & 3 & 0 & 0 & 12 & 0 & 0 \\
 0 & 0 & 11 & 0 & 0 & 38 & 0 \\
 4 & 0 & 0 & 32 & 0 & 0 & 92 \\
 0 & 12 & 0 & 0 & 75 & 0 & 0 \\
 0 & 0 & 38 & 0 & 0 & 160 & 0 \\
 10 & 0 & 0 & 92 & 0 & 0 & 313
\end{array}
\right)
\end{equation}

There are 16 generators, four $\epsilon \epsilon A A A$, three ${\rm Tr}(AB)$, five
${\rm Tr}(ABAB)$ and four
$\epsilon \epsilon B B B$. They transform as spin $\frac{3}{2}$, 1, 2, and $\frac{3}{2}$ of $SU(2)_R$, respectively. These generators are subject to 6 relations at order six and twelve
relations at order seven. Due to the non trivial F terms, 
there are no non factorizable baryons for $N=3$. For higher values of $N$,
the generators of the chiral ring include the basic determinants
$\epsilon \epsilon A A A ...$ and  $\epsilon \epsilon B B B ...$,
mesonic operators ${\rm Tr} (AB)^i$ for $i=1,...,N-1$ and non factorizable 
baryons. The latter appears for the first time at $N=4$; there are $N-3$ 
baryons of the form $\epsilon \epsilon (ABA) (A) (A) (A)$ as well as more
complicated determinants that appear for higher values of $N$.

\section{Conclusions}

In this paper we use two basic principles to construct the exact generating function which counts baryonic and mesonic BPS operators in the chiral ring of quiver gauge theories. Explicit formulas are given for the conifold theory and some subsets of the fields in it (these are termed ``half the conifold" and ``$\frac{3}{4}$ the conifold", respectively), and for the orbifold $\frac{\mathbb{C}^2}{Z_2}$. The first principle relies on the plethystic program which is presented in \cite{Benvenuti:2006qr} and discussed further in \cite{Feng:2007ur}. The second principle is based on \cite{Butti:2006au} and describes how to handle baryonic generating functions with specific divisors in the geometry. This is translated into generating functions for fixed baryonic numbers and hence leads to the full generating function for any number of D branes, $N$, and any baryonic charge, $B$, as well as the other charges in the system such as $R$ charges and flavor charges.

The results in the paper can be also interpreted as based on the semiclassical quantization of wrapped supersymmetric branes in an $AdS_5$ 
background dual to a strongly coupled gauged theory at large $N$ \cite{Butti:2006au}.  
It is interesting to note in this context that such countings 
seem to be valid also for weak coupling constant and small values of $N$, 
as similarly observed in the case of $\mathbb{C}^3$ \cite{Biswas:2006tj}. 

An important outcome of the study in this paper is the identification of the baryon number $B$ with the K\"ahler modulus. This identification repeats itself in both the conifold theory and in the orbifold $\frac{\mathbb{C}^2}{Z_2}$ and is expected to be true for any CY manifold. In fact the computation of generating functions reveals an important connection between gauge theories and their CY moduli spaces by providing the baryon number as the discrete area of two cycles. This is the quantum volume of the two cycle in a regime in which the string theory is strongly coupled and the gauge theory description reveals an underlying discrete structure. We expect this correspondence to produce a new line of research for strongly coupled string theories.

In this paper we considered in details the case of the conifold and the
simplest orbifold $\mathbb{C}^2/Z_2$ and we did not include $BPS$ fermionic 
operators. 
It would be interesting to extend the analysis done in this paper to the case 
of an arbitrary toric Calabi Yau with arbitrary number of baryonic symmetries
and to the case of fermionic operators. 
Another interesting direction of research would be the study of the large charges behavior of the baryonic generating functions and the construction of $BPS$ entropy functions for quiver gauge theories.
 
There are several other directions one can pursue in counting problems of supersymmetric gauge theories \cite{Sinha:2006sh, Noma:2006pe, Grant:2007ze, Nakayama:2007jy}. It would be very interesting to understand if there exist a relation between the type of counting presented in this paper and other counting problems such as the computations of microscopic degrees of freedom of black holes entropy and instanton partition functions. We plan to study these problems in future publications.

\section*{Acknowledgments}

D.~F.~ and A.~Z.~ would like to thank Agostino Butti and Bo Feng for valuable discussions.
A.~H.~ would like to thank Sergio Benvenuti, Freddy Cachazo, Dragomir {\DJ}okovi\'c, Bo Feng, Yang-Hui He, Christian Romelsberger, and David Vegh for enlightening discussions.
A.~H.~ would like to thank the physics department at the University of Cincinnati for kind hospitality during the final stages of this project.
D.~F.~ and A.~Z.~ are supported in part by INFN and MIUR under  
contract 2005-024045-004  and 2005-023102 and by 
the European Community's Human Potential Program
MRTN-CT-2004-005104.
Research at Perimeter Institute for Theoretical Physics is supported in part by the government of Canada through NSERC and by the Province of Ontario through MRI.


\bibliographystyle{JHEP}

\end{document}